\documentclass[a4paper,11pt,fleqn,final]{article}  % fleqn  draft  final

\usepackage{amssymb}
\usepackage{amsxtra}
\usepackage{mathcomp,mathrsfs}
\usepackage{amsmath,wasysym,textcomp,MnSymbol}		% marvosym
\usepackage{pifont}
\usepackage{booktabs}
\usepackage{fancyhdr,totpages}
\usepackage{rotating,psfrag}

\usepackage[applemac]{inputenc}
\usepackage[T1]{fontenc}
\usepackage{graphicx,psfrag,url} %,mdwlist}  \subsubsection
\usepackage[dvipsnames,usenames]{xcolor}
\graphicspath{{figures/}{../Slides/Sources/}}
\usepackage{pstricks,pstricks-add,pst-node,pst-plot}

\usepackage{enumitem}

\definecolor{BleuGris}{rgb}{0.392,0.475,0.635}

\definecolor{gris}{gray}{0.85}
\definecolor{gris2}{gray}{0.6}

\usepackage{titlesec}
\titleformat{\section}{\Large\bfseries}{\thesection}{1em}{} %{\hrule}
\titleformat{\subsection}{\large\bfseries}{\thesubsection}{1em}{} %{\hrule}
\titleformat{\subsubsection}{\large\it\centering}{\thesubsubsection}{1em}{} %{\hrule}

\usepackage[font={small,rm},nooneline]{caption}  %[2008/04/01]   %margin=0.5cm,
\DeclareCaptionFont{blue}{\color{BleuGris}}
\DeclareCaptionFormat{myformat}{#1#2\hrule\vspace*{0.1cm} #3}
\usepackage[round]{natbib}
\bibpunct{[}{]}{;}{a}{,}{,}
\fancypagestyle{frontmatter}{%
\fancyhf{} 
\fancyhead[L]{}
\fancyhead[C]{}

\fancyhead[R]{\textit{submitted to Earth and Planetary Science Letters}}
\fancyfoot[C]{ \thepage/\ref{TotPages}}}
\fancypagestyle{main}{%
\fancyhf{} 
\fancyhead[L]{}
\fancyhead[C]{}

\fancyhead[R]{\textit{submitted to Earth and Planetary Science Letters}}
\fancyfoot[C]{ \thepage/\ref{TotPages}}}
\headwidth   \textwidth 
\title{\bf Turbulent metal-silicate mixing, fragmentation,\\ and equilibration in magma oceans}
\author{Renaud Deguen$^{1,2}$\footnote{\tt renaud.deguen@univ-lyon1.fr}, Maylis Landeau$^{3,4}$ \& Peter Olson$^{3}$}%\\
\date{\small 
\footnotesize 
$^{1}$ Laboratoire de G\'eologie de Lyon, Universit\'e Claude Bernard Lyon 1,\\ Ecole Normale Sup\'erieure de Lyon, CNRS, France.\\
$^{2}$ Institut de M\'ecanique des Fluides de Toulouse, Universit\'e de Toulouse (INPT, UPS),\\ CNRS. All\'ee C. Soula, Toulouse, 31400, France.\\
$^{3}$ Department of Earth and Planetary Sciences, Johns Hopkins University, Baltimore, MD 21218, USA.\\
$^{4}$ Dynamique des Fluides G\'eologiques, Institut de Physique du Globe de Paris, Universit\'e Paris-Diderot, INSU/CNRS, 1 rue Jussieu, 75238, Paris cedex 05, France.
}
\begin{document}

\newcommand{\R}[1]{\textcolor{black}{#1}}

\renewcommand{\thefootnote}{\fnsymbol{footnote}}
\renewcommand{\labelitemi}{$\filledtriangleright$} 

\pagestyle{main}

\maketitle

\begin{abstract}

Much of the Earth was built by high-energy impacts of \R{planetesimals and embryos}, many of these impactors already differentiated, with metallic cores of their own. 
Geochemical data provide critical information on the timing of accretion and the prevailing physical conditions, but their interpretation depends critically on the  degree of metal-silicate chemical equilibration during core-mantle differentiation, which is poorly constrained. 
Efficient equilibration requires that the large volumes of iron derived from impactor cores mix with molten silicates down to scales small enough to allow fast metal-silicate mass transfer. 
Here we use fluid dynamics  experiments  to show that large metal blobs falling in a magma ocean  mix with the molten silicate through turbulent entrainment,  
with fragmentation into droplets eventually resulting from  the entrainment process.
In our experiments, fragmentation of the dense fluid occurs  after falling a distance equal to 3-4 times its  initial diameter, at which point a sizable volume of ambient fluid has already been entrained and mixed with the dense falling fluid.
Contrary to what has usually been assumed, we demonstrate that fragmentation of the metallic phase into droplets may not be required for efficient equilibration:  turbulent  mixing, by drastically increasing the metal-silicate interfacial area, may result in fast equilibration even before fragmentation.
\R{%Small impactors, in the sense that ...
Efficient re-equilibration is predicted for impactors of size small compared to the magma ocean depth.
In contrast, a much smaller re-equilibration degree is predicted in the case of large impacts for which the impactor core diameter approaches the magma ocean thickness.  
%A higher re-equilibration level may be reached if the impactor core 
%... as seen in some SPH simulations of the Moon-forming giant impact.
%The case of the largest impacts, including the Moon-forming giant impacts, are probably close to the boundary  
}

\end{abstract}

%\begin{keyword}
%%% keywords here, in the form: keyword \sep keyword
%
%core formation \sep magma ocean \sep fragmentation \sep turbulent mixing \sep chemical equilibration
%
%%% MSC codes here, in the form: \MSC code \sep code
%%% or \MSC[2008] code \sep code (2000 is the default)
%
%\end{keyword}



\section{Introduction}

The formation of Earth's core  produced  chemical and isotopic fractionations which have been used to constrain the timing of differentiation \citep{Yin2002,Kleine2002} and the physical conditions \citep{Wood2006,Corgne2008,Siebert2011,Rubie2011} that prevailed early in Earth's history.
Hafnium-Tungsten (Hf-W)  systematics in particular  provide  constraints on the timing of accretion, but their interpretation   depends critically on the degree to which the metal portion of the impactors equilibrates isotopically with Earth's mantle silicates \citep{Halliday2004,Kleine2004,Nimmo2010,Rudge2010}.  
Assuming full equilibration after each impact, Hf-W chronometry implies an  accretion timescale of about $10$ My assuming an exponentially decreasing accretion rate  \citep{Yin2002,Rudge2010}, whereas relaxing this assumption can increase this timescale by several tens of My, or even render it indeterminate \citep{Rudge2010}.

Partial equilibration is usually modeled by assuming that  a fraction $k$ of the metal phase \R{delivered by each impact} re-equilibrates with the whole mantle, the remaining metal fraction $1-k$ reaching the Earth's core without chemical interaction with the mantle \citep{Halliday2004,Kleine2004,Nimmo2010,Rudge2010}. 
%\R{Clearly, the assumption that the whole mantle is equilibrated}
However,  the  compositional transfer between metal and silicate also depends on the quantity of silicates the metal phase equilibrates with. For example, the amount of radiogenic Tungsten extracted from the silicates by the metal will be insignificant if the volume of interacting silicate is small. 
We thus define a more general measure of equilibration, the equilibration efficiency $\mathscr{E}_i$,  as   the total mass of element $i$ exchanged between metal and silicates normalized by its maximum possible value, had all the metal re-equilibrated with an infinitely larger silicate reservoir. 
If a fraction $k$  of the metal phase equilibrates with a mass of silicates equal to $\Delta$ times the mass of equilibrated metal, the equilibration efficiency of an element $i$ with a metal/silicate partition coefficient $D_i$ is, from mass balances,
\begin{equation}
\mathscr{E}_i = \frac{k}{1+D_i/\Delta}  \label{EquilibrationEfficiency}	
\end{equation}
(see Appendix \ref{AppendixEquilibrationEfficiency}), with the metal dilution $\Delta$ defined as
 \begin{equation}
 \Delta = \frac{\mbox{mass of equilibrated silicates}}{\mbox{mass of equilibrated metal}}. \label{MetalDilution}
 \end{equation}
$\mathscr{E}_i$ approaches $k$ when $\Delta \gg D_i$, which is the usual assumption of disequilibrium core formation models.
Importantly, $\mathscr{E}_i$ is element-dependent, with efficient equilibration of an element $i$ requiring a metal dilution $\Delta$ similar or larger than its distribution coefficient. 
Tungsten, for example, had \R{a mean distribution coefficient around} $D_W\simeq 30$ \R{during Earth's differentiation}, so that equilibration is efficient only if the metal mixes and equilibrates with more than about 30 times its mass of silicates \R{on average}.

Previous disequilibrium geochemical models assuming infinite dilution can be  corrected for the effect of finite metal dilution by  substituting $\mathscr{E}_i$ in place of $k$  (\R{as demonstrated in} \ref{AppendixEquilibrationEfficiency}), which means that previously determined constraints on $k$  actually apply to $\mathscr{E}_i$. %, \R{which can provides constraints on both $k$ and $\Delta$}.	% would
In particular, Hf-W systematics imply that the  Tungsten equilibration efficiency $\mathscr{E}_W$ must have been larger than about $0.36$ on average during Earth's accretion \citep{Rudge2010}, which requires that on average $k\geq 0.36$ and \R{$\Delta/D_W \geq 0.56$}. % $\Delta\geq 17$. 
\R{In practice, the distribution coefficient of W may have changed by several order of magnitude in the course of Earth's accretion due to possible changes in oxygen fugacity  \citep{Cottrell2009,Rubie2011}, and this makes the process of obtaining constraints on metal-silicate mixing from Hf-W systematics a non-trivial matter.
% and this would have to be taken into account for deriving a ... lower bound on a mean value of $\Delta$.
As an illustration, assuming an average $D_W$ around 30 \citep{Rudge2010} implies an average metal dilution $\Delta$ larger than about $17$,  which argues for significant metal-silicate mixing.
}
\R{Though Hf-W systematics can provide a lower bound on the degree of metal-silicate mixing and equilibration, its use as a core-formation chronometer is still hampered by the lack of stronger constraints on the degree of equilibration: there is an inverse trade-off between the assumed degree of re-equilibration and the Hf-W accretion timescale, which even 
becomes unbounded when  $\mathscr{E}_i$ approaches it's lower acceptable bound (0.36 according to \cite{Rudge2010}).  
%The Hf-W accretion timescale becomes unbounded 
Additional constraints on metal-silicate equilibration are  needed to properly interpret the data.
}
%The Hf-W accretion timescale becomes unbounded when  $\mathscr{E}_i$ approaches 0.36 \citep{Rudge2010}, so that additional constraints on metal-silicate equilibration are  needed to properly interpret the data.

During accretion, dissipation of the gravitational and kinetic energies associated with large impacts inevitably results in  widespread melting  \citep{Melosh1990,Tonks93,Pierazzo1997}, implying that part of the separation of the core-forming metal phase from the silicates occurred in low-viscosity magma oceans.
Under these conditions, efficient chemical equilibration would be expected if the Earth had formed through the accretion of undifferentiated bodies with the metal phase already finely dispersed within a silicate matrix.
However, it is now recognized that much of the Earth  was accreted from already differentiated bodies with sizes ranging from a few tens of kilometers in diameter to objects the size of Mars \citep{Yoshino2003,Baker2005,Bottke06,Ricard2009}.
It is usually assumed that efficient chemical equilibration between the cores of these impactors and the proto-Earth's mantle requires fragmentation of the metal down to scales of $1$ cm to 1 m where efficient metal-silicate chemical equilibration can occur \citep{Stevenson90,Karato1997,Rubie2003,Ulvrova2011}, implying a scale reduction by a factor of $10^4-10^8$. 
Smooth Particle Hydrodynamics (SPH) simulations of the Moon-forming impact  suggest some degree of disruption of the impactor core into 100-1000 km sized iron blobs \citep{Canup2004}, but the current resolution of these models  is too coarse to give any information about  smaller scale  mixing and fragmentation. 
Hence the fate of these large iron blobs, while critical for the interpretation of geochemical data, remains uncertain.

\section{Non-dimensional parameters}
\label{NonDimensionalParameters}

We consider the evolution of an iron blob, which can be either the core of an impactor or a fragment of an impactor core, falling in a magma ocean. 
Its dynamics are characterized by the following set of non-dimensional numbers~:
\begin{align*}
Re &= \R{ \frac{w\, d}{\nu_m}},\  &W\!e &= \frac{\rho_m\, w^2 d}{\sigma}, \  &Bo &= \frac{\Delta \rho\, g\, d^2}{\sigma}, \\ 
%Re &=  \frac{\rho_m\,w\, d}{\eta_m},\  &W\!e &= \frac{\rho_m\, w^2 d}{\sigma}, \  &Bo &= \frac{\Delta \rho\, g\, d^2}{\sigma}, \\ 
M &= \frac{w}{c},   &\mathsf{P}&= \frac{\rho_m}{\rho_s},  & \mathsf{H}&= \frac{\eta_m}{\eta_s},  
\end{align*}
%\R{Reynolds ecrit en fonction de viscosite cinematique ??}
where $w$ and $d$ are the velocity and diameter of the falling metal volume,  $\rho$ is density, $\eta$  the dynamic viscosity, \R{$\nu=\eta/\rho$ the kinematic viscosity}, $g$ the acceleration of gravity, $\sigma$ the iron-silicate interfacial tension, and $c$ the sound wave velocity  in the dominant phase.
Subscripts "m" and "s" refer to  metal and silicate, respectively, and $\Delta \rho=\rho_m-\rho_s$. 
The Reynolds number $Re$ compares the magnitude of inertia  to viscous forces,  the Weber and Bond numbers, $W\!e$ and $Bo$, are measures of the relative importances of \R{inertia and buoyancy} to interfacial tension at the lengthscale $d$, and the Mach number $M$ compares the  velocity of the flow to the sound wave velocity. 
%\R{Another useful dimensionless number is the Ohnesorge number $Oh=$}
\R{A list of the symbols used in the main text and appendices is given in table \ref{Symbols}.}

Typical values for these parameters for a metal blob 100 km in diameter falling in a magma ocean with an initial velocity of $1$ km.s$^{-1}$ are $Re\sim 10^{14}$, $Bo\sim 10^{14}$, $W\!e\sim 10^{14}$, with  $\mathsf{P} \simeq 2$ and $\mathsf{H} \sim 0.1- 1$.
Note that $Re$, $W\!e$, $Bo$ and $M$ are all time-dependent.

The huge value of $Re$ implies that the flow must have been extremely turbulent. 
The Weber and Bond numbers are large as well, which implies that interfacial tension effects were unimportant except at the smallest scales of the flow \citep{Dahl2010,Deguen2011c}.
%$M$ is typically larger than 1 (supersonic flow), \R{maybe up to $\sim 5$}, just after the impact and decreases with time as the metal decelerates, \R{implying that compressibility effects are important}.
%\R{The flow is typically supersonic, with the Mach number $M$ maybe up to $\sim 5$ just after the impact, and decreasing with time as the metal decelerates, implying that compressibility effects are important.}
\R{The Mach number $M$ may be up to $\sim 5$ just after the impact (with an impact velocity $\sim 15$ km.s$^{-1}$ and a speed of sound  $\sim3$ km.s$^{-1}$), and then decreases with time as the metal decelerates. The flow is typically supersonic, implying that compressibility effects are important.}

\section{Turbulent entrainment} 

\begin{figure}[t]
\begin{center}
\includegraphics[width=0.9\linewidth]{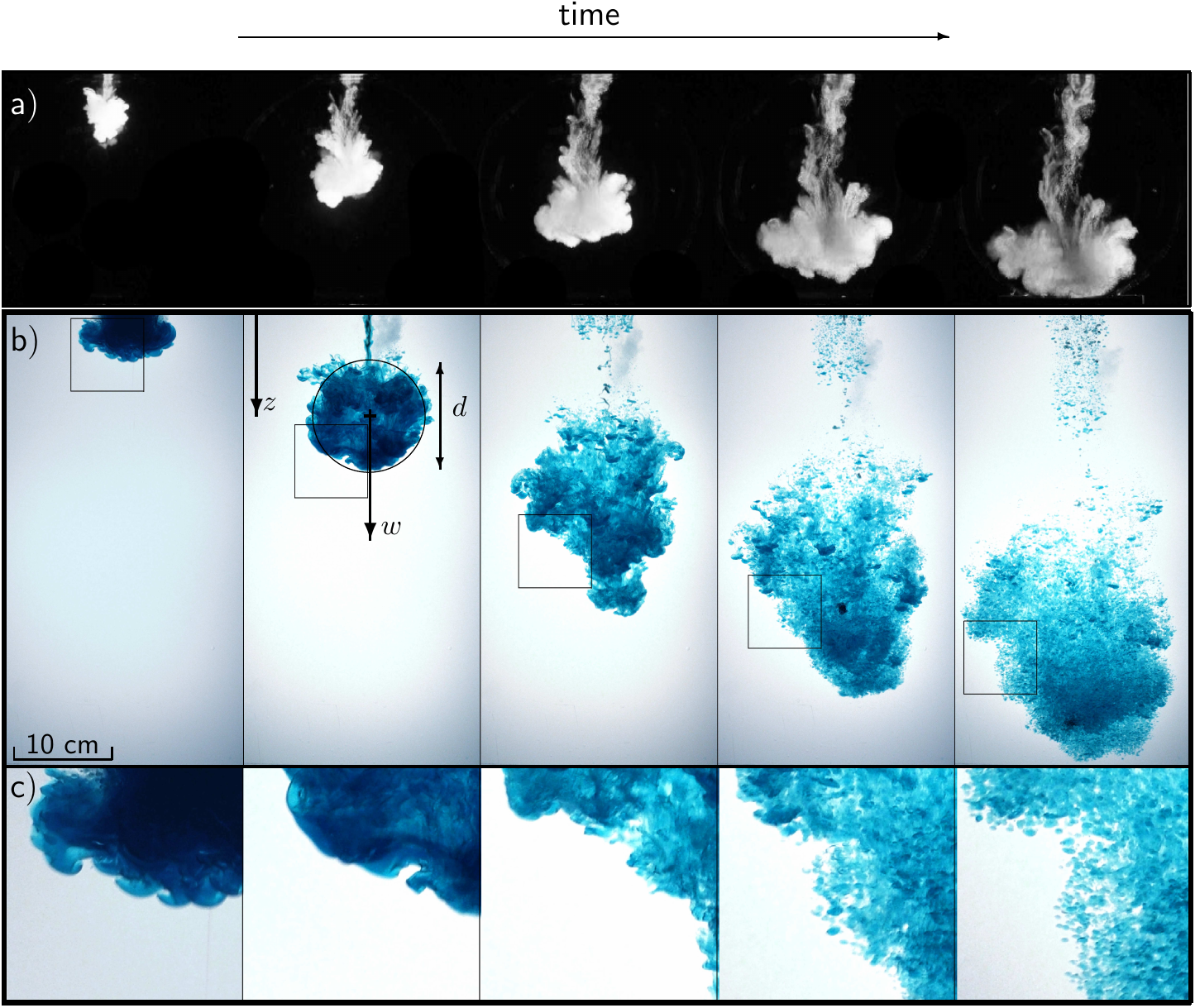}
\caption{{Analog fluid dynamics experiments on metal-silicate mixing and fragmentation.}
\textbf{a)} Growth of a negatively buoyant thermal by turbulent entrainment at $Re=2\times10^3$. Here the buoyancy of the falling fluid is due to very fine dense particles in suspension (modified from   \cite{Deguen2011c}). 
A small amount of fluorescent dye (fluoresceine) is added to the particle-laden fluid, which appears white in the pictures.
\textbf{b)} Fragmentation of a volume of aqueous solution of NaI salt (dyed in blue) released in silicone oil, at $W\!e=3\times 10^3$, $Re=2\times10^4$, $\mathsf{P}=1.9$, $\mathsf{H}=2.1$.
Fragmentation of the aqueous volume into droplets occurs between the third and fourth snapshots.
\textbf{c})  Close-ups corresponding to the squares in \textbf{b)}.
Small scale Rayleigh-Taylor  instabilities are apparent in the first close-up.   
\label{snapshots}
}
\end{center}
\end{figure}

% $W\!e$ and $Bo$
Given the extreme values of the Weber and Bond numbers, it is appropriate to first consider  the limiting case of miscible fluids, for which $W\!e$ and $Bo$ are formally infinite. 
Numerous  experimental and theoretical studies  have shown that the evolution of a turbulent buoyant fluid falling or rising under the action of gravity - what is called a \textit{turbulent thermal} in fluid mechanics - is governed by turbulent entrainment of ambient fluid \citep{Batchelor1954,Morton1956,Turner1986}.
As an illustration, Fig. \ref{snapshots}a shows snapshots from an experiment in which a  volume of a dense solution is released into a larger volume of pure water. 
A small amount of fluorescent dye has been added to the solution.
The volume of dyed fluid is seen to increase as it falls, which indicates that the negatively buoyant fluid entrains and incorporates ambient fluid,  resulting in its  gradual dilution \citep{Batchelor1954,Morton1956}.

This effect is quantified using the \textit{entrainment hypothesis} of  \cite{Morton1956}, which states that the rate of entrainment of ambient fluid is proportional to the mean velocity of the buoyant turbulent fluid, and predicts that the radius $r=d/2$ of the buoyant fluid evolves as
\begin{equation}
r = r_0 + \alpha\, z,  \label{ScalingRadius}    
\end{equation}
where $\alpha$ is the entrainment coefficient and $r_0=d_0/2$ the initial radius of the dense blob.
The velocity of the mixture can be calculated from the equations of conservation of momentum and mass (\ref{AppendixTurbulentEntrainment}), a general expression being given in Eq. \eqref{Solution_w2}.
The velocity law \eqref{Solution_w2} has a useful large-$z$ asymptote given by 
\begin{equation}
 w = \left( \frac{r_0^3 g}{2\, \alpha^3} \frac{\Delta \rho}{\rho_s} \right)^{1/2} \left( 1+ K + \frac{3}{16} \frac{C_d}{\alpha}\right)^{-1/2} \frac{1}{ z}  \label{AsymptoteVelocity},
\end{equation}
where $C_d$ is the drag coefficient, and $K$ the coefficient of added mass, which accounts for the momentum imparted to the surrounding fluid.
These  laws have been verified in a wide variety of physical settings, from laboratory experiments using thermally or compositionally buoyant fluids to large scale geophysical flows including explosive volcanic plumes \citep{Terada2007,Yamamoto2008}, underwater gas plumes \citep{Bettelini1993},  and atmospheric convective bursts (known as \textit{thermals} - hence the name - by sailplane pilots \citep{woodward1959}).

Turbulent entrainment results from a combination of \textit{engulfment} of ambient fluid by large scale, inviscid eddies, which draws large volumes of surrounding fluid into the turbulent region, and \textit{nibbling}, which denotes small scale viscous processes (vorticity diffusion)  \citep{Turner1986,Mathew2002,WESTERWEEL2009}.
The rate  at which the ambient fluid is entrained is thought to be controlled by large scale process \citep{Brown1974,Turner1986}, while nibbling is responsible for eventually imparting vorticity to the entrained fluid. 
The entrainment coefficient appears to be independent of $Re$ \citep{Turner1969}, which is consistent with the rate of turbulent entrainment being controlled by the largest inviscid eddies rather by the small scale viscous effects. 
In two-fluids systems we would expect that these \R{large-scale} eddies remain unaffected by interfacial tension if the Weber number is large enough, in which case turbulent entrainment should still occur, at a rate similar to the case of miscible fluids. 
We  argue here that the concept of turbulent entrainment is indeed also applicable to immiscible fluids like molten metal and silicate, provided $Re$ and $W\!e$ are large. 
This is demonstrated below in a series of experiments with two immiscible fluids. 

\section{Experimental set-up}

Molten silicate is modeled by a low viscosity silicone oil (density $\rho_s=820$ kg.m$^{-3}$, viscosity $\eta_s=1$ mPa~s) enclosed in a 25.5~cm $\times$ 25.5~cm $\times$ 47~cm container. 
A volume of NaI aqueous solution (density $\rho_m=1580$ kg.m$^{-3}$, viscosity $\eta_m=2$ mPa~s), representing a metal blob falling  into a magma ocean, is held in a vertically oriented tube whose lower extremity is sealed using a thin latex diaphragm, which is ruptured at the beginning of the experiment.
\R{Tube diameters from 1.28 cm to 7.62 cm have been used, with an aspect ratio (height of fluid in the tube/diameter of the tube)  kept to 1 in all  experiments. }
A surfactant (Triton X-100) is added to the NaI solution, lowering the interfacial tension of the silicone oil/NaI solution system to about $5$ mJ~m$^{-2}$.
A small amount of Na$_2$S$_2$O$_3$ is added to the NaI solution to avoid a yellowish coloration of the solution.
In experiments where induced fluorescence is used to image cross-sections (Fig. \ref{CrossSection}), we use a concentration of the NaI solution for which the refractive index of the NaI solution matches that of the silicone oil, which is necessary to avoid optical distortions. 
At this concentration, its density is $\rho_m=1260$ kg.m$^{-3}$.
The exact values of the densities, viscosities and interfacial tension are measured before each series of experiments.
The experiments are recorded with a color video camera at 24 frames per second.
Using a pixel intensity threshold method, we estimate on each video frame the location of the center of mass $z$ of the oil/NaI solution mixture and the apparent area $A$ of the mixture, from which its equivalent radius is estimated as $r=\sqrt{A/\pi}$.

The dense fluid is released from rest and its vertical velocity is  set by the conversion of its gravitational potential energy into kinetic energy, which implies that the vertical velocity initially scales as $w\sim \sqrt{({\Delta\rho}/{\rho_m}) g\, r}$. 
Using this scaling for $w$ implies that $W\!e\sim Bo$, using the equivalent diameter of the NaI solution volume as the length scale.
The  Weber and Reynolds \R{numbers} that characterize the experiments are defined using as a velocity scale the vertical velocity of the dense fluid after it has travelled a distance equal to its initial diameter. 
With this definition, we found that  $W\!e\simeq 0.43\, Bo$ in our experiments.
Our choice of experimental fluids plus the use of a surfactant to reduce the interfacial tension allows us to reach values of $Re$ larger than $10^4$ and $W\!e$ up to $ 3\times 10^3$, making our experiments far more dynamically similar to planetary accretion than current numerical simulations \citep{Ichikawa2010,samuel2012}.

\R{%??? MAYBE WE NEED SOMETHING LIKE THIS HERE ???  
We have explored a wide range of parameters, with density ratios $\mathsf{P}$ from sligtly larger than 1 to about 2, and Reynolds and Weber numbers ranging from moderate values to around $10^4$ and $3\times 10^3$, respectively.
We focus here on the experiments we performed at the largest Reynolds and Weber numbers and a density ratio similar to that of the metal-silicate system, which are the most relevant to the core-mantle differentiation problem. 
More details about the all set of experiments will be found in a companion paper \citep{Landeau2013}.
}

\section{Experimental validation of the turbulent entrainment model}
\label{SectionExperimentalEntrainment}

Snapshots from an experiment with \R{Bond number} $Bo=6.9\times10^3$, \R{Weber number} $W\!e=3\times 10^3$, \R{Reynolds number} $Re=2\times10^4$, \R{density ratio} $\mathsf{P}=1.9$, and \R{viscosity ratio} $\mathsf{H}=2.1$ are shown in Fig. \ref{snapshots}b and c. 
After release, the dense fluid (dyed in blue) undergoes small scale Rayleigh-Taylor instabilities (apparent on the first snapshot) which, together with shear induced by the global motion of the fluid, generate turbulence.
The volume of the falling fluid  increases with time much like the miscible fluids case shown in Fig. \ref{snapshots}a, indicating that entrainment is occurring in spite of immiscibility.

\begin{figure}[ht]
\begin{center}
\includegraphics[width=0.44\linewidth]{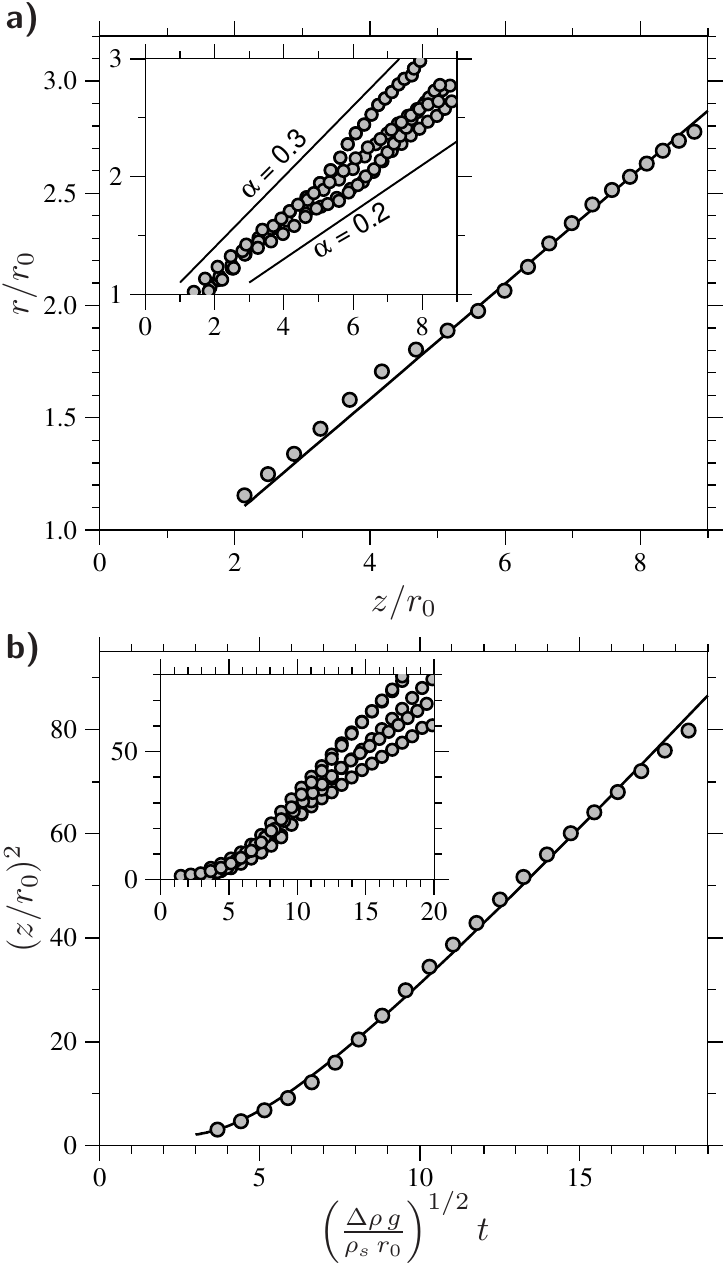}
\end{center}
\vspace*{-0.7cm}
\caption{{Time evolution of the mean radius and position of the falling fluid,} 
in experiments where a volume of  NaI solution is released into silicone oil ($\mathsf{P}=1.9$).
\textbf{a)} Mean radius $r$ (normalized by $r_0$) of the aqueous solution/oil mixture as a function of the position $z$ (normalized by $r_0$) of its center of mass. 
\textbf{b)} Square  of the normalized position $z$ of the center of mass of the aqueous solution/oil mixture as a function of time (normalized by $\left({\Delta\rho\,g}/{\rho_s\,r_0}\right)^{-1/2}$).
The results of one experiment are compared with the predictions of our model based on the entrainment assumption  shown with black lines in \textbf{a)} and \textbf{b)}. 
For this experiment, the model [Eqs. \eqref{mass_conservation3}, \eqref{momentum_conservation3} and \eqref{InitialConditions}] best fits the data with $\alpha=0.26$, a drag coefficient $C_d=0.53$, and a virtual mass coefficient $K=0.5$ (see \ref{AppendixTurbulentEntrainment}  for details on the model).
The experimental results shown in the inserts illustrate the natural variability seen in our experiments, with $\alpha$ varying between 0.2 and 0.3.
\label{fig_r_z}}
\end{figure}

Fig. \ref{fig_r_z} shows that the equivalent radius of the NaI solution-silicone oil mixture increases linearly with the distance travelled,  in agreement with the turbulent entrainment model predictions (Eq. \eqref{ScalingRadius}). 
The entrainment coefficient $\alpha$ is in the range 0.2-0.3 in our experiments, similar to turbulent thermals in miscible fluids \citep{Morton1956,Turner1969}, which suggests that we have indeed reached a regime for which the large scales of the flow are unaffected by interfacial tension effects.

The predicted descent trajectory  also compares favorably with the experimental results.
Once integrated in time, the asymptotic velocity law Eq. \eqref{AsymptoteVelocity} yields  
\begin{equation}
\left(\frac{z}{r_0}\right)^2 = \left(\frac{2\,\Delta \rho\, g}{\alpha^3 \rho_s\, r_0}\right)^{1/2} \left( 1+ K + \frac{3}{16} \frac{C_d}{\alpha}\right)^{-1/2} t. \label{z2}  
\end{equation}
Fig. \ref{fig_r_z}b shows that  after a short acceleration phase the experiments agree well with the prediction of Eq. \eqref{z2} that $z^2 \propto t$, although there is some variability in the magnitude of the slope.
The full evolution of our experiments can be explained by  the model described in \ref{AppendixTurbulentEntrainment}.
Although the drag and virtual mass coefficients are uncertain,  the  model (black curves in Fig. \ref{fig_r_z}) fits very well the experimental measurements for reasonable values of these coefficients, with the observed variability  in our experiments attributable to imperfect control of initial conditions plus natural variability inherent in turbulent flows.

The  agreement between our experiments and the entrainment prediction strongly supports our contention that the turbulent entrainment concept can be applied to immiscible fluids when $W\!e$ and $Re$ are large, and offers a simple way [Eq. \eqref{ScalingRadius}, \eqref{AsymptoteVelocity}, and  \ref{AppendixTurbulentEntrainment}] to model the evolution of large metal masses in a magma ocean.
In particular, the linear increase of the buoyant mixture radius provides a measure of metal-silicate mixing, with the metal dilution [Eq. \eqref{MetalDilution}] given by
\begin{equation}
\Delta = \frac{\rho_s}{\rho_m}\left[\left( 1 + \alpha \frac{z}{r_0}  \right)^{3} - 1\right]. \label{Phi}  
\end{equation}

%We have so far ignored the  effects of compressibility on the entrainment process, which are negligible in our experiments but may be significant if the post-impact flow is supersonic or nearly supersonic.
%The fact that the flow velocity is similar to the sound velocity has an  important qualitative consequence for the structure of the flow: the finite speed of sound introduces a time delay in the transmission of pressure signals from one point to another, which makes it impossible for large turbulent eddies to remain coherent when the local Mach number (based on the eddy velocity scale) is of order one \citep{breidenthal1992,Freund2000,Pantano2002}. 
%Because the rate of entrainment is thought to be controlled by the process of engulfment of ambient fluid by large scale eddies \citep{Brown1974,Turner1986,Mathew2002}, mixing is expected to decrease when ${M}\rightarrow 1$.
%Experiments on compressible turbulent jets and mixing layers show that the entrainment rate indeed decreases significantly with increasing $M$, before saturating at a value about five times smaller than for incompressible flows \citep{Brown1974,Freund2000} when $M\gtrsim 1$.

\section{Fragmentation}

Fig. \ref{snapshots}b-c reveals that the dense NaI solution entrains and incorporates silicone oil \textit{before} it fragments into droplets. 
Fragmentation occurs relatively late in the descent process (between the third and fourth pictures in the experiment shown in Fig. \ref{snapshots}b-c), at a time when a sizable volume of ambient fluid has already been entrained. 
Droplets  appear in a single global fragmentation event, which is at variance with previously suggested "cascade" processes, in which a succession of fragmentation events lead to the final stable drop size \citep{Rubie2003,samuel2012}, and   "erosion" processes, in which metal-silicate mixing occurs predominantly on the boundary with the ambient fluid \citep{Dahl2010}. 

\begin{figure}[t]
\begin{center}
\includegraphics[width=0.7\linewidth]{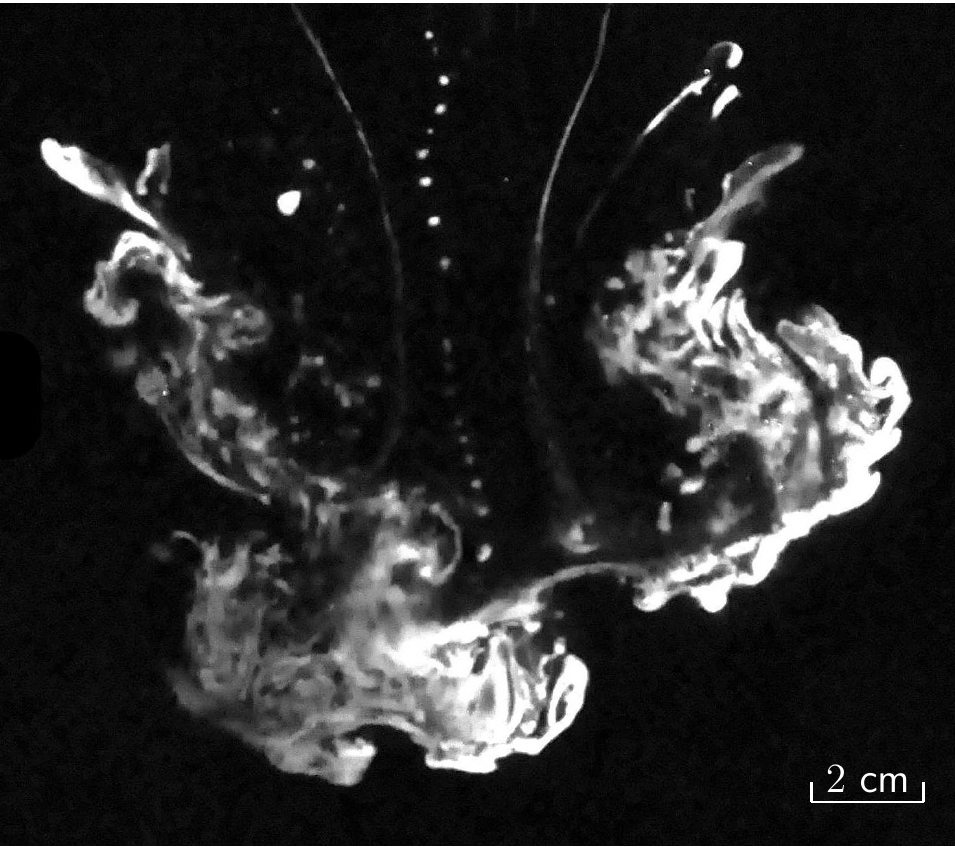}
\end{center}
\vspace*{-0.7cm}
\caption{{Cross-section of the NaI solution/silicone oil mixture at a distance $\sim 2\, d_0$ from the origin.}
The experiment is illuminated with a thin light sheet exciting a fluorescent dye (Rhodamine B) added to the NaI solution, which appears white in the picture.
In this experiment $Bo=4.6\times10^3$, $W\!e=2\times10^3$, $\mathsf{P}=1.54$, $\mathsf{H} = 2.1$, and $Re=2\times10^4$.
\label{CrossSection}
}
\end{figure}

Adding a small amount of fluorescent dye to the NaI solution and illuminating the experiment with a thin light sheet reveals cross-sections of the NaI solution/silicone oil mixture, one example being shown in Fig. \ref{CrossSection}.
Small scale mixing of the phases is evident in this picture, demonstrating that oil has  been entrained into the NaI solution and that the two phases are already intimately mixed  \textit{before} fragmentation  occurs.
This striking observation suggests  that fragmentation is a consequence of mixing associated with turbulent entrainment of the ambient fluid, with fragmentation into drops ultimately resulting from small scale instabilities, plausibly capillary instabilities developed on filaments stretched by the turbulent flow  \citep{villermaux2004,Shinjo2010}.

In all our experiments in this turbulent  regime, fragmentation into drops is observed to occur after the dense liquid falls a distance equal to 3 to 4 times its initial diameter, with no clear trend observed in the explored range of parameters. 
At this point the volume fraction of the dense fluid in the mixture is of order 5-10 \%.
It is possible that the fragmentation distance becomes independent of $Re$ and $W\!e$ when these two numbers are large, but the maximum value of $W\!e$ obtained in our experiments (3000) is only 6 times larger than its observed critical value for this turbulent regime ($\sim500$), making the explored range of $W\!e$ too small to test this possibility.

\section{Chemical equilibration before fragmentation - a fractal model \label{SectionFractal}}

Fragmentation of the metal phase into drops is an important facet of the problem of metal-silicate interactions, because drop formation is an efficient way of increasing the interfacial area between metal and silicate, thus enhancing chemical transfer and equilibration.
However, it may  not be necessary for chemical equilibration.
The small scale mixing  observed in our experiments (Fig. \ref{CrossSection}) results in a highly convoluted interface, which should drastically decrease the timescale of equilibration with the entrained silicate. 

\begin{figure}[t]
\begin{center}
\includegraphics[width=0.7\linewidth]{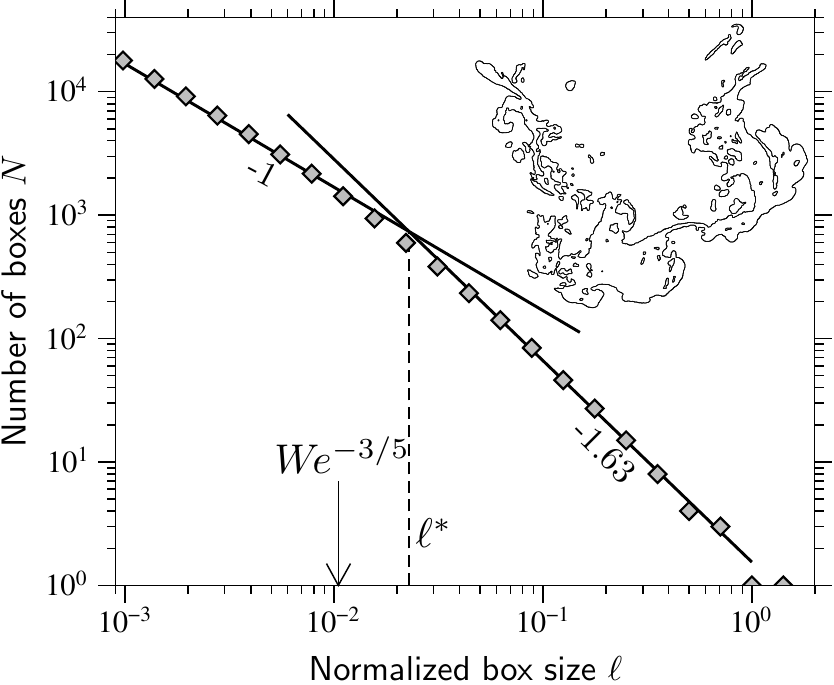}
\end{center}
\vspace*{-0.7cm}
\caption{{The fractal dimension of the oil/aqueous solution interface shown in Fig. \ref{CrossSection}}, determined using a  box counting algorithm.
Shown here is the number $N(\ell)$ of square boxes of size $\ell$ required to cover the oil/aqueous solution interface as a function of the box size $\ell$. \R{Here the box size $\ell$} is normalized by the size of the smallest box fully enclosing the interface.
The slope of the resulting curve is  $1-D$, where $D$ is the fractal dimension of the 3D interface.
A slope of $-1$ is expected for a non-fractal surface, as found here for small $\ell$.
The interface is fractal at scales above $\ell^*\simeq2.3\times10^{-2}$, which is of the same order of magnitude as $W\!e^{-{3}/{5}}\simeq1.05\times10^{-2}$.
Fitting the data for $\ell>\ell^*$ (thick black line) gives a slope of $-1.63\pm0.03$ ($\pm1\, \sigma$), which implies a fractal dimension of $2.63\pm0.03$, slightly smaller than $D=8/3=2.67$.
\label{fig_fractal}
}
\end{figure}

To illustrate this point, we consider a model of metal-silicate equilibration prior to drop formation based on the observation that the interface separating the two fluids has a fractal nature once turbulence is well-developed. 
Theory \citep{Mandelbrot1975,Constantin1991,Constantin1994} and experiments \citep{Sreenivasan1989,Constantin1991} show that isosurfaces of transported quantities (composition, temperature) in well-developed turbulent flows are fractal -- a consequence of the self-similarity of the turbulent flow -- with a fractal dimension  predicted to be $D=8/3$ for homogeneous turbulence with Kolmogorov scaling. 

It is to be expected that the interface between immiscible fluids in a turbulent flow shares this property over the range of scales in which interfacial tension is unimportant.
Experimental support for this assumption is given in  Fig. \ref{fig_fractal}, where the interface between the oil and aqueous solution is shown to have a fractal nature with a fractal dimension  at scales larger than a cut-off length $\ell^*$. 
For miscible fluids,  \cite{Sreenivasan1989} assumed that the inner cut-off length is the Kolmogorov scale for isovorticity surfaces, and the Batchelor scale for isocompositional surfaces for high Schmidt number fluids.
For a surface separating two immiscible fluids, we expect that the inner cut-off length will be the largest of the Kolmogorov scale $\ell_K=d\, Re^{-3/4}$ and the scale $\ell_\sigma = d\, W\!e^{-3/5}$ at which interfacial tension balances local dynamic pressure fluctuations estimated assuming a Kolmogorov cascade [\citep{Kolmogorov1949,Hinze1955}, and see section \ref{DropSize} for more details].
Typically $\ell_\sigma \gg \ell_K$, and we expect that $\ell^* \sim \ell_\sigma$. 
In our experiments, $\ell^*$ and $\ell_\sigma$ are numerically close (within a factor of 2, Fig. \ref{fig_fractal}) and the measured fractal dimension  is only slightly smaller than the theoretical value of $8/3$.
Note that the observed fractal nature of the interface is indicative of self-similarity in the flow, and that the measured fractal dimension is consistent with Kolmogorov type turbulence and a $k^{-5/3}$ kinetic energy spectrum.

Assuming that the metal-silicate interface has a fractal nature  offers a  convenient way of estimating its   area $A_T$, which according to fractal geometry is $A_T = A_0 (\ell^*/d)^{2-D}$, where $A_0=\pi d^2$ is the area measured at the scale $d$. 
Using $\ell^*\sim \ell_\sigma$, the predicted surface area is $A_T\sim A_0 W\!e^{\frac{3}{5}(D-2)}$. With $D=8/3$ and $W\!e=10^{14}$, this implies an increase in interfacial area by five orders of magnitude. 
A timescale for chemical equilibration, $\tau_\mathrm{eq}$, can then be found by coupling the estimate for $A_T$ with a local scaling for turbulent mass flux at the metal-silicate interface.

\begin{figure}[t]
\begin{center}
\includegraphics[width=0.7\linewidth]{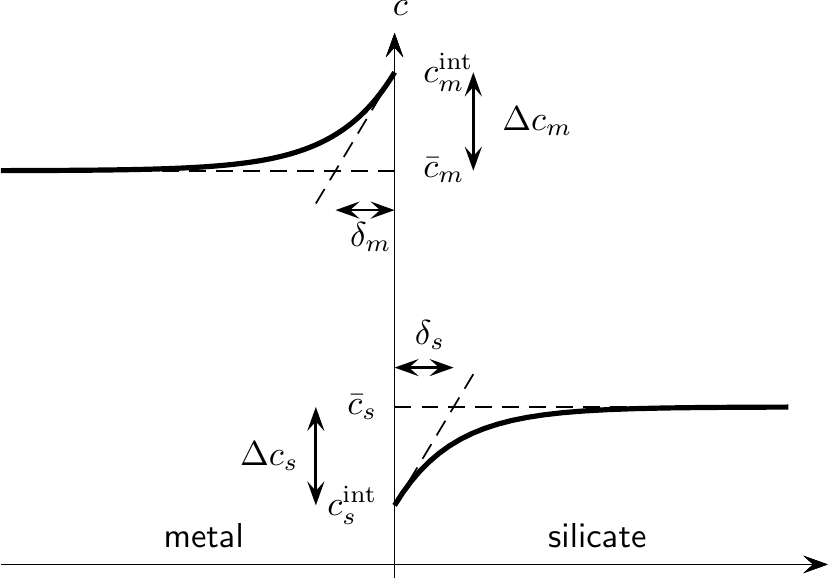}
\end{center}
\vspace*{-0.7cm}
\caption{A sketch of the composition profiles in the vicinity of  the metal-silicate interface.
The situation depicted here is that of a siderophile element in excess in the silicate phase.
\label{FigCProfile}
}
\end{figure}

We denote by $\kappa_c$ the diffusivity of the chemical element of interest. The Schmidt number $S\!c=\nu/\kappa_c$, where $\nu$ is the kinematic viscosity, is assumed to be large in both phases.
Fig. \ref{FigCProfile} shows a sketch of the composition profiles in the vicinity of  the metal-silicate interface, with definitions of the main variables.
Thermodynamic equilibrium is assumed at the metal/silicate interface, so that the concentrations by mass $c_m^\mathrm{int}$ and $c_s^\mathrm{int}$ at the interface are linked by the partition coefficient $D_i=c_m^\mathrm{int}/c_s^\mathrm{int}$, but the bulk compositions $\bar c_m$ and $\bar c_s$ are out of thermodynamic equilibrium, $\textit{i.e.}$ $\bar c_m/\bar c_s \neq D_i$.
The resulting compositional boundary layers have thicknesses $\delta_{m,s}$, and we denote by $\Delta c_{m,s}$ the composition difference across the  boundary layers.
The local diffusive compositional flux across the interface  scales as $\kappa_c \Delta c/\delta$ and the total mass flux $F\!_c$ is
\begin{equation}
F\!_c \sim \rho_m A_T \kappa_c^m \frac{\Delta c_m}{\delta_m} \sim \rho_s A_T \kappa_c^s \frac{\Delta c_s}{\delta_s}.  \label{InterfaceFluxScaling}
\end{equation}
Continuity of the mass flux across the interface implies that \R{the ratio $ \gamma_{m/s}$ of $\Delta c_m$ to $\Delta c_s$ is}
\begin{equation}
 \gamma_{m/s} = \frac{\Delta c_m}{\Delta c_s} = \frac{\rho_s}{\rho_m} \frac{\kappa_c^s}{\kappa_c^m}\frac{\delta_m}{\delta_s}.  \label{gamma_ms}
\end{equation}
We now  relate the compositional jumps $\Delta c_s$ and $\Delta c_m$ to the  mean composition $\bar c_m$ and $\bar c_s$ of the metal and silicate phases.
Using Eq. \eqref{gamma_ms} together with the assumption of local thermodynamic equilibrium ($D_i=c_m^\mathrm{int}/c_s^\mathrm{int}$), we obtain  the following expressions for $\Delta c_s$ and $\Delta c_m$~:
\begin{equation}
\Delta c_s = - \frac{\bar c_m - D_i \bar c_s}{ \gamma_{m/s}+D_i},\quad \Delta c_m = - \gamma_{m/s} \frac{\bar c_m - D_i \bar c_s}{ \gamma_{m/s}+D_i}.
\end{equation}

Using $(\pi/6) \bar \rho\, d^3$ for the mass of the metal-silicate mixture, 
the evolution of composition in the metal and silicate phases are given by
\begin{align}
\phi \frac{\pi}{6} \bar \rho\, d^3 \frac{d \bar c_m}{dt} &= - F\!_c,   \label{C_conserv_m} \\
(1-\phi) \frac{\pi}{6} \bar \rho\, d^3 \frac{d \bar c_s}{dt} &= F\!_c,  \label{C_conserv_s}
\end{align}
where $\phi$ is the mass fraction of the metal phase in the mixture.
Combining Eqs. \eqref{C_conserv_m} and \eqref{C_conserv_s} and using the metal dilution $\Delta=(1-\phi)/\phi$, we obtain
\begin{equation}
\frac{d}{dt}\ln\left(  \bar c_m - D_i \bar c_s \right) = - \frac{(1+\Delta)(D_i+\Delta)}{\Delta(D_i+\gamma_{m/s})}  \frac{\rho_s}{\bar \rho}\,  \frac{6 \kappa_c^s}{d\, \delta_s}\, W\!e^{\frac{3}{5}(D-2)} , \label{EquilEq}
\end{equation}
from which we obtain an equilibration timescale $\tau_\mathrm{eq}$ \R{of order of magnitude} given by
\begin{equation}
\tau_\mathrm{eq} =f(\Delta,D_i,\gamma_{m/s})\,  \frac{\bar \rho}{\rho_s}\,\frac{d\,\delta_s}{\kappa_c^s}\, W\!e^{-\frac{3}{5}(D-2)},  \label{tau_eq_1}
\end{equation}
where the factor 6 in Eq. \eqref{EquilEq} has been omitted, \R{on the basis that this expression for $\tau_\mathrm{eq}$ is based on an order of magnitude estimate of the flux across the interface [Eq. \eqref{InterfaceFluxScaling}], in which an unknown -- presumably $\mathcal{O}(1)$ -- factor has already been omitted. 
%The equilibration timescale $\tau_\mathrm{eq}$ given by Eq. \eqref{tau_eq_1} should really be seen as an order of magnitude estimate. 
In Eq. \eqref{tau_eq_1}, the function $f(\Delta,D_i,\gamma_{m/s})$ is given by}
\begin{equation}
f(\Delta,D_i,\gamma_{m/s}) = \frac{\Delta (D_i+\gamma_{m/s})}{(1+\Delta)(D_i+\Delta)}.		
\end{equation}
The function $f$ is $\mathcal{O}(1)$ for intermediate values of $\Delta$ (with a maximum always smaller than 1), but $f\rightarrow 0$ if $\Delta$ is small compared to $\min(1,D_i)$ or large compared to $\max(1,D_i)$.

We  now estimate the boundary layers thicknesses $\delta$ in the metal and silicate phases (the subscript $m$ and $s$ will be omitted in what follows, with the understanding that the analysis applies to both phases).
Denoting by $\ell$ the smallest scale of the flow in the  vicinity of the interface, then the smallest scale $\delta$ of the compositional field is found by balancing the strain rate  at scale $\ell$ with the diffusion rate at the scale $\delta$, \textit{i.e.} $u_{\ell} / \ell \sim \kappa_c/{\delta}^2$.
Assuming  a Kolmogorov type velocity spectrum, the velocity at scale $\ell$ is $u_{\ell}\sim w (\ell/d)^{\frac{1}{3}}$, where $w$ is the large scale velocity. With these assumptions, we obtain
\begin{equation}
\delta = d\, {S\!c}^{-\frac{1}{2}} {Re}^{-\frac{1}{2}} \left( \frac{\ell}{d} \right)^{\frac{1}{3}} . \label{delta_c_2}
\end{equation}
At this stage, further progress \R{requires}  assumptions on the small scale structure of the turbulence in the vicinity of the metal-silicate interface :
\begin{enumerate}
\item 
If we  assume that the turbulence structure is not affected by the presence of the interface and interfacial tension effects, then $\ell$ should be the Kolmogorov scale. 
Eq. \eqref{delta_c_2} with $\ell=\ell_K=d Re^{-3/4}$ gives
\begin{equation}
\delta = d\, S\!c^{-\frac{1}{2}} Re^{-3/4}, 
\end{equation}
which is the Batchelor scale $\ell_B$. 
With this estimate for $\delta$, we obtain 
\begin{equation}
\gamma_{m/s}  = \left( \frac{\rho_s}{\rho_m} \right)^{5/4} \left( \frac{\kappa_c^s}{\kappa_c^m} \right)^{1/2} \left( \frac{\eta_s}{\eta_m} \right)^{1/4}
\end{equation}
and an equilibration timescale 
\begin{equation}
\tau_\mathrm{eq} =f(\Delta,D_i,\gamma_{m/s})\,  \frac{\bar \rho}{\rho_s}\,\frac{d^2}{\kappa_c^s}\, S\!c^{-1/2}  Re^{-3/4} W\!e^{-\frac{3}{5}(D-2)}. \label{EqTimescaleK}
\end{equation}

\item 
Alternatively, one might argue that the turbulent motion in the  vicinity of the interface is damped by interfacial tension at scales smaller than $\ell_\sigma$.
In this case the smallest scale of the flow is $\ell_\sigma \sim d\, W\!e^{-3/5}$ and the boundary layer thickness is
\begin{equation}
\delta = d\, S\!c^{-\frac{1}{2}} Re^{-\frac{1}{2}}W\!e^{-\frac{1}{5}},
\end{equation}
which gives
\begin{equation}
\gamma_{m/s}  = \left(\frac{\rho_s}{\rho_m} \right)^{6/5} \left(\frac{\kappa_c^s}{\kappa_c^m} \right)^{1/2}
\end{equation}
and an equilibration timescale
\begin{equation}
\tau_\mathrm{eq} =f(\Delta,D_i,\gamma_{m/s})\,  \frac{\bar \rho}{\rho_s}\,\frac{d^2}{\kappa_c^s}\, S\!c^{-\frac{1}{2}} Re^{-\frac{1}{2}}W\!e^{-\frac{3}{5}D+1}. \label{EqTimescaleSigma}
\end{equation}
\end{enumerate}

\begin{figure}[t]
\begin{center}
\includegraphics[width=0.7\linewidth]{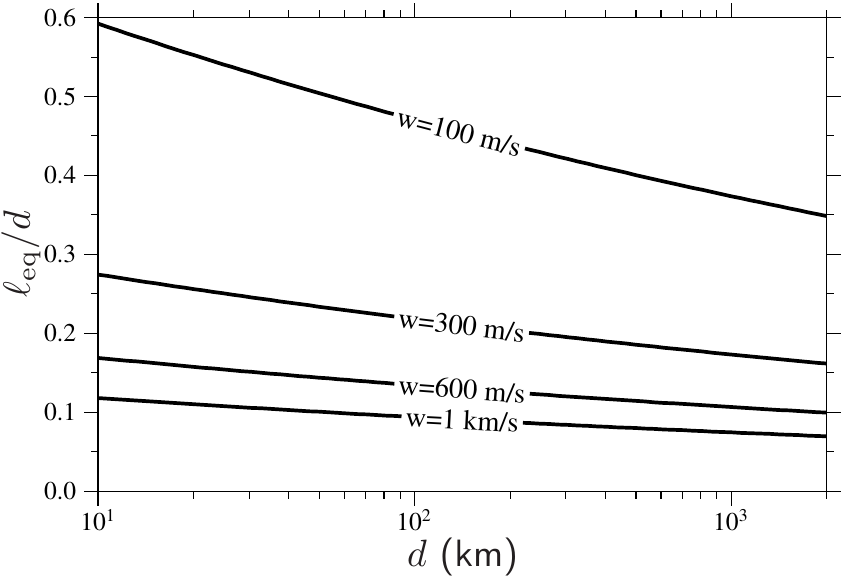}
\end{center}
\vspace*{-0.7cm}
\caption{Equilibration distance $\ell_\mathrm{eq}/d$ as  functions of the metal-silicate mixture diameter $d$, for $w=100,300,600$ and 1000 m.s$^{-1}$, calculated using Eq. \eqref{lEqTimescaleSigma} with $f(\Delta,D_i)=0.5$, $\kappa_c^s\sim10^{-8}$ m.s$^{-1}$, $\sigma=1$~J.m$^{-2}$ and $\rho_s=3500$ kg.m$^{-3}$.
\label{Fig_l_eq}		
}
\end{figure}

Choosing between the two models Eqs. \eqref{EqTimescaleK} or \eqref{EqTimescaleSigma} would require detailed measurement of the small scale structure of the flow, or alternatively,  measurements of a tracer concentration in both phases, which are beyond the scope of our current experimental set-up.
We therefore choose the more conservative estimate of the equilibration timescale  Eq. \eqref{EqTimescaleSigma} which assumes that turbulent motions in the vicinity of the interface are damped at scales smaller than $\ell_\sigma$. 
For comparison, the model assuming no effect of the interface on the turbulence structure would yield an equilibration timescale a factor $W\!e^{1/5}Re^{-1/4}$ smaller (typically a factor of 5 or more smaller).

With $\rho_s/\rho_m \simeq 0.5$,  assuming that $\kappa_c^s$ and $\kappa_c^m$ are of the same order of magnitude implies that $\gamma_{m/s}=\mathcal{O}(1)$.
Since it only appears in $f(\Delta,D_i,\gamma_{m/s})$ as a sum with $D_i$ which is $\gg 1$ for siderophile elements, the exact value of $\gamma_{m/s}$ should be of little importance.
The factor ${\bar \rho}/{\rho_s}$ is also $\mathcal{O}(1)$, and ignoring it as well in Eq. \eqref{EqTimescaleSigma} yields the simplified equilibration timescale  % used in the paper :
\begin{equation}
\tau_\mathrm{eq} =f(\Delta,D_i)\, \frac{d^2}{\kappa_c^s}\,S\!c^{-\frac{1}{2}} Re^{-\frac{1}{2}} W\!e^{-\frac{3}{5}D+1}. \label{EqTimescaleSigma2}
\end{equation}
%From Eq. \eqref{EqTimescaleSigma2}, the equilibration distance $\ell_\mathrm{eq} = w\,  \tau_\mathrm{eq}$ is
From Eq. \eqref{EqTimescaleSigma2}, \R{we obtain an  equilibration distance $\ell_\mathrm{eq} = w\,  \tau_\mathrm{eq}$ given by}
\begin{equation}
\ell_\mathrm{eq} = f(\Delta,D_i)\,  d\, S\!c^{\frac{1}{2}} Re^{\frac{1}{2}} W\!e^{-\frac{3}{5}D+1}, \label{lEqTimescaleSigma}
\end{equation}
\R{which is the distance travelled by the metal phase during the time $\tau_\mathrm{eq}$ required for equilibration of the metal with the silicate it has mixed with. 
%Equilibration is efficient
Metal and silicates equilibrate if the equilibration distance $\ell_\mathrm{eq}$ is smaller than the magma ocean depth.}
Fig. \ref{Fig_l_eq} shows $\ell_\mathrm{eq}$ as a function of $d$ for various values of $w$ between $100$ m.s$^{-1}$ and 1 km.s$^{-1}$, calculated with $f(\Delta,D_i)=0.5$, $\kappa_c^s\sim10^{-8}$ m.s$^{-1}$, $\sigma = 1$ J.m$^{-2}$ and $\rho_s=3500$ kg.m$^{-3}$.
The equilibration distance is always a fraction of the metal-silicate mixture diameter, and is usually smaller than plausible magma ocean depths.

\section{Prediction for the stable drop size after fragmentation \label{DropSize}}

After fragmentation, the metal-silicate equilibration timescale depends mostly on the resulting fragments size \citep{Karato1997,Rubie2003,Ulvrova2011}. 
\R{In this section, we propose a scaling for the drop size of the metal phase after fragmentation, as well as a justification for the scaling used in the previous section for the cut-off length scale $\ell^*$ above which the interface (before fragmentation) is fractal.}
%  \R{introduced in section \ref{SectionFractal}}
In a fully turbulent flow, the stable drop size $d_d$ after fragmentation, as well as the cut-off length scale $\ell^*$ before fragmentation, are expected to depend only on the dissipation rate $\epsilon$, the interfacial tension $\sigma$,  the densities and viscosities of both phases, and the metal volume fraction~:
\begin{equation}
\R{(d_d,}\ell^*\R{)} = \mathcal{F}_1(\epsilon,\sigma,\rho_m,\rho_s,\nu_s,\nu_m,\phi).
\end{equation}
Using the Vashy-Buckingham theorem, we find that $\ell^*$ must be the solution of an equation of the form
\begin{equation}
\mathcal{F}_2\left[\mathsf{P},\mathsf{H},\phi,\frac{\ell^*}{\ell_K},\frac{\ell^*}{\ell_\sigma} \right] = 0.
\end{equation}
where we have introduced  two length scales,
\begin{equation}
\ell_K = \left( \frac{\nu_s^3}{\epsilon} \right)^{1/4}, \quad \ell_\sigma = \left( \frac{\sigma}{\rho_s} \right)^{3/5} \epsilon^{-2/5}.
\end{equation}
$\ell_K$ is the Kolmogorov scale, at which turbulent kinetic energy is dissipated into heat by the action of viscous forces; $\ell_\sigma$ can be shown to be the length scale at which interfacial tension (Laplace pressure) balances turbulent pressure fluctuations and  stresses if a Kolmogorov type turbulence is assumed  \citep{Kolmogorov1949,Hinze1955}.
With $\epsilon \sim w^3/r$  \citep{Tennekes1972}, 
\begin{equation}
\ell_K \sim  Re^{-3/4} d, \quad \ell_\sigma \sim  W\!e^{-3/5} d.
\end{equation}

Two end-member cases are possible, depending on the relative values of $\ell_K$ and $\ell_\sigma$. 
Let us first compare the magnitude of the viscous stress and Laplace pressure at a given scale $\ell$. 
Assuming a Kolmogorov type turbulence cascade, the velocity fluctuations $u_{\ell}$ at  scale $\ell$ is $u_{\ell} \sim w \left( \ell/d\right)^{1/3}$.
Using this estimate for $u_{\ell}$, we find that the ratio of the viscous stress to the Laplace pressure at the scale $\ell$ is 
\begin{equation}
\frac{\text{Viscous stress at scale } \ell}{\text{Laplace pressure at scale } \ell} \sim \frac{\eta_s u_{\ell}/\ell}{\sigma/\ell} \sim \left( \frac{\ell_K}{\ell_\sigma} \right)^{4/3} \left(\frac{\ell}{\ell_\sigma} \right)^{1/3}.   \label{ViscousLaplaceRatio}
\end{equation}
Two options are possible :
\begin{enumerate}
\item First, if $\ell_K \gg \ell_\sigma$, all the energy input is dissipated at the Kolmogorov scale, at which scale the ratio of viscous stress and Laplace pressure is $\sim (\ell_K/\ell_\sigma)^{5/3} \gg1$ according to Eq. \eqref{ViscousLaplaceRatio}.
In this case  interfacial tension is unimportant, and \R{$d_d$ and} $\ell^*$  scales as
\begin{equation}
\R{(d_d,}\ell^*\R{)} = \mathcal{F}_3\!\left( \mathsf{P},\mathsf{H},\phi\right) \left( \frac{\nu_s^3}{\epsilon} \right)^{1/4} \sim \mathcal{F}_3\!\left( \mathsf{P},\mathsf{H},\phi\right) d\, Re^{-3/4}.
\end{equation}
\item Alternatively, if $\ell_K \ll \ell_\sigma$, then interfacial tension  balances turbulent pressure and stress fluctuations at the scale $\ell_\sigma$, with further smaller scale deformation of the interface inhibited by the interfacial tension.
According to Eq. \eqref{ViscousLaplaceRatio}, the ratio of viscous stress and Laplace pressure is $\sim (\ell_K/\ell_\sigma)^{4/3} \ll1$ at this scale, which implies that viscous effects are unimportant.
As a consequence, the stable drop size does not depend on the viscosity of either phase, nor on the viscosity ratio $\mathsf{H}$, and  thus \R{the drop size and}  the cut-off length scale  follow a  scaling law of the form:
\begin{equation}
\R{(d_d,}\ell^*\R{)} =\mathcal{F}_4\!\left( \mathsf{P},\phi\right)   \left( \frac{\sigma}{\rho_s} \right)^{3/5} \epsilon^{-2/5} \sim \mathcal{F}_4\!\left( \mathsf{P},\phi\right) d\, W\!e^{-3/5}.	\label{ScalingWe}
\end{equation}
\end{enumerate}

The ratio ${\ell_K}/{\ell_\sigma} \sim W\!e^{3/5} Re^{-3/4}$ following an impact is found to be typically smaller than $10^{-2}$, which suggests that the drop size or cut-off length will be set by interfacial tension rather than viscosity, and will obey the scaling given by Eq. \eqref{ScalingWe}.
When $\phi$ is small, its effect  should be negligible, as indeed observed in experiments with dilute dispersions  \citep{Hinze1955,Chen1967}.

\begin{figure}[t]
\begin{center}
\includegraphics[width=0.7\linewidth]{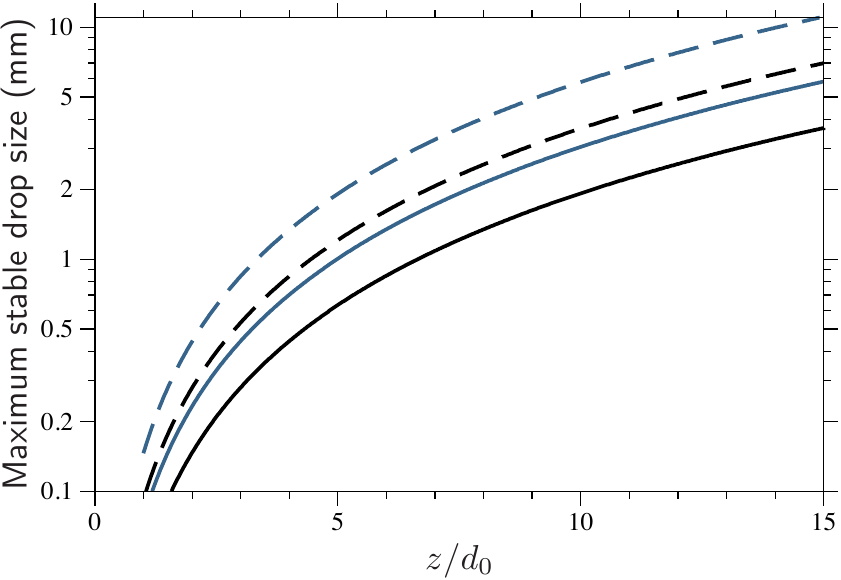}
\caption{Maximum stable drop size after fragmentation according to Eq. \eqref{StableDropSize}, as a function of the distance travelled (normalized by the initial metal blob diameter $d_0$), for metal blobs with initial diameter 100 km (blue curves) and 1000 km (black curves) with $f=0.5$ (solid curves) and $f=0.1$ (dashed curves).
Assumed parameters values are : $\alpha=0.25$, $K + {3\, C_d}/{16\,\alpha} = 1$, $\Delta \rho=4000$ kg~m$^{-3}$, $\rho_s=3500$ kg~m$^{-3}$, $g=5$ m~s$^{-2}$, $\sigma=1$ J~m$^{-2}$.
\label{FigStableDropSize}
}
\end{center}
\end{figure}

From analysis of  \cite{Clay1940}'s data,  \cite{Hinze1955} found that the maximum drop size $d_\mathrm{max}$ in a turbulent flow with $\ell_\sigma \gg \ell_K$ is given by 
\begin{equation}
d_\mathrm{max} \simeq 0.725\, \left( \frac{\sigma}{\rho_s} \right)^{3/5} \epsilon^{-2/5}. \label{DropSizeHinze}
\end{equation}
Effects of changing the density ratio was not  investigated in this study, which focused on fluids with density ratios $\mathsf{P}\simeq 1$. 
Theory  \citep{Levich1962} and experiments  \citep{Hesketh1987} argue for a dependence on the density ratio of the form $d_\mathrm{max}\propto \mathsf{P}^{-1/5}$.
For the metal-silicate system, which has $\mathsf{P}\simeq 2$, this would predict a maximum drop size about $13$ \% smaller than what Eq. \eqref{DropSizeHinze} predicts,  a  minor discrepancy in light of the other uncertainties.

Clearly, the size of the drops produced by fragmentation of the metal blob must depend on the details of the fragmentation mechanism, which are not elucidated yet, and the drop size just after fragmentation does not have to match the prediction of Eq. \eqref{DropSizeHinze} (although a similar scaling is expected).  
Nevertheless, Eq. \eqref{DropSizeHinze} should  give a reasonable upper bound for the fragment size, since it predicts that larger drops would be disrupted by turbulent dynamic pressure fluctuations.

In a system in statistical steady state, the dissipation rate $\epsilon$ must  equal  the total energy input in the system $e_\mathrm{in}$, which here is the rate of work of the buoyancy forces.
However, since the metal-silicate mixture is not in statistical steady state (it can be shown using the self-similar regime velocity (Eq. \eqref{AsymptoteVelocity}) that the total kinetic energy of the system evolves with time), dissipation does not equal the rate of energy input, but is some fraction $f$ of the work done by the buoyancy forces.
The rate of work of the buoyancy forces,
\begin{equation}
e_\mathrm{in} =\bar\phi\, \frac{\Delta \rho}{\bar \rho}\, g\, w,
\end{equation}
tends towards
\begin{equation}
e_\mathrm{in} = 2 \left[ 1+K + \frac{3}{16}\frac{C_d}{\alpha}  \right]^{-1/2}  \frac{\rho_s}{\bar \rho} \left[\frac{\Delta\rho\,g\, r_0^{1/3}}{2 \alpha^3 \rho_s}\right]^{3/2}\left(\frac{r_0}{z}\right)^4, \label{BuoyancyForcesWork}
\end{equation}
in the self-similar regime, for which $w$ is given by Eq. \eqref{AsymptoteVelocity}.
Using Eq. \eqref{BuoyancyForcesWork} for $e_\mathrm{in}$ and writing the dissipation as $\epsilon = f\, e_\mathrm{in}$, we find that 
\begin{equation}
\frac{d_\mathrm{max}}{d_0} \simeq \frac{3}{f^{2/5}}  \left[ 1+K + \frac{3}{16}\frac{C_d}{\alpha}  \right]^{1/5}  \left(   \frac{\bar \rho}{\rho_s}\right)^{2/5}  \frac{ \alpha^{9/5}}{ Bo_0^{3/5}} \left(\frac{z}{d_0}\right)^{8/5}  \label{StableDropSize}
\end{equation}
when the mixture has reached the self-similar regime. 
Here $Bo_0=\Delta \rho\, g\,d_0^2/\sigma$.
The value of $f$ is difficult to estimate precisely, but shouldn't be much smaller than 1.
Fig. \ref{FigStableDropSize} shows $d_\mathrm{max}$ from Eq. \eqref{StableDropSize} for metal blobs with initial diameter 100 km (blue curves) and 1000 km (black curves) with $f=0.5$ (solid curves) and $f=0.1$ (dashed curves), and $\alpha=0.25$. Smaller values of $\alpha$ would result in smaller drop sizes.
Eq. \eqref{StableDropSize} predicts submillimeter-to-centimeter maximum stable drop sizes, which is small enough to ensure fast re-equilibration  with the surrounding silicates \citep{Karato1997,Rubie2003,Ulvrova2011}.

\R{\section{Discussion}}

%\begin{figure}
%\psset{arrowsize=2pt 3,unit=0.07\linewidth}
%\begin{center}
%\begin{pspicture}(2,0)(16,10)  
%   \psline[linecolor=black,arrowsize=6pt]{<->}(3,10.5)(3,1)(15.2,1)
%   \uput[-90](15.2,1){$\log k$}
%   \uput[180](3,10.5){$\log E(k)$}
%   
%   \psline[showpoints=false,linearc=1,linewidth=1.5pt,linecolor=black]{}(3.5,5)(5,10.6)(13,5.2667)(14.5,1.5)
%   \uput[180](10.1,8.6){\textcolor{black}{-5/3}}
%   \psline[linecolor=black,arrowsize=2pt,linestyle=dashed]{}(5.5,1)(5.5,9.9)
%   \uput[-90](5.5,1){$1/d$}
%   \psline[linecolor=black,arrowsize=2pt,linestyle=dashed]{}(13,1)(13,5)
%   \uput[-90](13,1){$1/\ell_K$}
%   \uput[-90](13.7,0.2){Dissipative }
%   \uput[-90](13.7,-0.3){scale}
%
%   \psline[linecolor=black,arrowsize=5pt,linewidth=0.5pt]{<->}(5.5,10.5)(13,10.5)
%   \uput[90](9.5,10.5){Inertial range}
%   
%   \psline[linecolor=black,arrowsize=2pt,linestyle=dashed]{}(11.4,1)(11.4,6.2)
%   \uput[-90](11.4,1){$1/\ell_\sigma$}
%   \uput[-90](11,0.2){\textcolor{black}{Capillary}}
%   \uput[-90](11,-0.3){\textcolor{black}{scale}}
%
%
% \end{pspicture}
% \end{center}
%\caption{}
%\end{figure}

%\R{\subsection{Limits}}
\R{\subsection{On the relevance of our experiments for the core formation problem} \label{Limits}}

\R{
Uncertainties about the applicability of our results to metal-silicate
mixing and fragmentation in magma oceans are due mostly
to the fact that our experiments are still very far from the impact
conditions with $Re$, $We$ and $Bo$ up to ten orders of magnitude smaller 
than during Earth accretion. The main obstacles to improving 
experimental as well as numerical approaches stem from the three dimensional,
turbulent nature of the flow at these extreme parameters. 
Direct numerical simulations of turbulent flows at such high Re are prohibitively 
expensive. For example, the cost\footnote{In a turbulent flow, the smallest scale which has to be resolved is  the Kolmogorov scale $\ell_K \sim d\, Re^{-3/4}$.
This therefore requires $\sim Re^{3/4}$ grid points in each direction, or $\sim Re^{9/4}$ grid points for a 3D simulations. 
The typical timescale corresponding to the Kolmogorov scale is $\tau_K=Re^{-1/2}d/w$, which means that $\sim Re^{1/2}$ timesteps are needed for a simulation time corresponding to $d/w$.
The total cost therefore scales as $\sim Re^{9/4}\times Re^{1/2} \sim Re^{11/4}$.
} of a direct numerical simulation resolving the Kolmogorov scale goes as $\sim Re^{11/4}$, which implies that increasing $Re$ by a factor of 10 multiplies the cost by about 500.}

\R{
Accordingly, a legitimate question is : how close to the dynamical conditions of accretion do we need to go ?
In boundary free turbulent flows involving fully miscible fluids, it is observed that there is no qualitative change of the flow associated with increasing $Re$ once turbulence is "fully-developed", $\textit{i.e.}$ once there is a range of length scales (the inertial range) for which viscosity effects are negligible \citep[\textit{e.g.}][]{mungal1989}. %does not play any role.
% scale separation between the largest scale of the flow and the scale at which 
% One  answer, based on analogy with turbulent flow involving miscible fluids, is that we need to have
%Decreasing further the viscosity ()
%There are good reasons to think that 
Increasing $Re$ further increases the gap between the integral scale (the largest scale of the flow, here the diameter of the blob) and the Kolmogorov scale at which viscous dissipation occurs, but does not change the slope of the kinetic energy spectrum. 
%In particular, t
The entrainment coefficient in turbulent thermals appears to be independent of $Re$ \citep[\textit{e.g.}][]{Turner1969} once $Re \gtrsim 10^3$, which is consistent with the rate of turbulent entrainment being controlled by the largest, inviscid eddies \citep{Turner1986}. 
}

\R{
In immiscible fluid systems like metal-silicate, we should expect that a similar asymptotic regime is reached once there is a separation of scales between the integral scale of the flow  and the Kolmogorov ($d\,Re^{-3/4}$) and capillary  ($d\, W\!e^{-3/5}$) scales, so that there is a range of scales for which viscosity and interfacial tension do not play any role.
% (the diameter of the blob)
%In immiscible fluid systems like metal-silicate, this requires a separation of scales between the integral scale of the flow  and the Kolmogorov and capillary scales ($d\, We^{-3/5}$) so that there is a range of scales for which viscosity and interfacial tension don't play any role. 
Increasing further $Re$ and $W\!e$ will  increase the ratio between the largest and smallest scales of the flow, but should not change the phenomenology, nor the slope of the kinetic energy spectrum in the inertial range.
%In turbulent thermals, the entrainment coefficient appears to be independent of $Re$, which is consistent with the fact that the rate of turbulent entrainment is thought to be controlled by the largest, inviscid eddies \citep{Turner1986}. 
%In two-fluid systems, a second requirement, based on the order of magnitude analysis of section \ref{NonDimensionalParameters}, is that interfacial tension effects have to be negligible at the large scale of the flow.
By analogy with the miscible fluid case, the entrainment coefficient in the immiscible fluids case should not depend on $W\!e$ and $Re$ once these numbers are large enough.
}

\R{
While it is difficult to demonstrate without heavier instrumentation and actual velocity measurements that our experiments have indeed reached a large $Re$, large $W\!e$ asymptotic regime, 
%... but there are a number of observations which are consistent with such regime being reached in our experiments :
%... but 
there are a number of observations which are consistent with our experiments being at least close to such regime:
(i) the measured coefficient of entrainment is similar to that measured in miscible turbulent thermals, consistent with the entrainment rate being independent of the interfacial tension; 
(ii) the observed fractal nature of the interface is indicative of self-similarity in the flow, and the measured fractal dimension is consistent with a $k^{-5/3}$ spectrum;
(iii) the cut-off length observed in cross-sections of the mixture, which presumably corresponds to the capillary scale, is more than a decade smaller than the diameter of the NaI-silicon oil mixture (40 times smaller in Fig. \ref{fig_fractal}).
}

\R{
Together, these observations support our claim that the entrainment model, and the entrainment coefficient value of $\simeq 0.25$ which we observe, should indeed apply to larger values of $Re$ and $W\!e$. 
%..., which 
%These ... argues ... robust and we believe that ... should indeed apply to larger values of $Re$ and $W\!e$. 
%In contrast, 
However, there is one more point which needs to be discussed :  compressibility effects, which are absent in our experiments (the Mach number $M$ is  $\sim 10^{-4}$), may be significant in the flow following an impact, which can often be supersonic. % or nearly supersonic.
This is probably the most severe limitation of our experiments. 
The fact that the flow velocity is similar to the sound velocity has an  important qualitative consequence for the structure of the flow: the finite speed of sound introduces a time delay in the transmission of pressure signals from one point to another, which makes impossible for large turbulent eddies to remain coherent when the local Mach number (based on the eddy velocity scale) is of order one or larger \citep{breidenthal1992,Freund2000,Pantano2002}. 
Because the rate of entrainment is thought to be controlled by the process of engulfment of ambient fluid by large scale eddies \citep{Brown1974,Turner1986,Mathew2002}, mixing is expected to decrease when ${M}$ approaches $1$.
Experiments on compressible turbulent jets and mixing layers show that the entrainment rate indeed decreases significantly with increasing $M$, before saturating at a value about five times smaller than for incompressible flows \citep{Brown1974,Freund2000} when $M\gtrsim 1$. 
An entrainment coefficient several times smaller than the $0.25\pm 0.05$ value of our experiments might therefore be expected when the Mach number of the metal-silicate mixture is $\mathcal{O}(1)$.
}

\R{\subsection{Comparison with previous work}}

\R{The reduction of a large metal blob to a drop size has been 
investigated by \cite{Dahl2010} and \cite{samuel2012}.}
%Our results are overall consistent with \cite{Dahl2010}. 
%We find that interfacial tension and viscosity are unimportant for the large scale component of the flow and for predicting large scale metal-silicate mixing. 
%\textcolor{red}
{Two different scenarios have been considered by \cite{Dahl2010}. 
In a first model, metal-silicate mixing is assumed to occur through gradual erosion of the metal blobs by small scale Rayleigh-Taylor instabilities. 
%In a first model, \cite{Dahl2010} assumed that metal-silicate mixing occurs through gradual erosion of the metal blobs by small scale Rayleigh-Taylor instabilities. 
%their second model is similar to the one we 
%Their model predicts  efficient metal-silicate mixing for relatively small metal blobs only, less than 10 km in diameter, larger blobs reaching the protoplanet core without significant chemical interactions with the surrounding silicates. 
The model predicts that only relatively small metal blobs (less than 10 km in diameter) efficiently mix with silicates, larger blobs reaching the core of the growing planet without significant chemical interactions with the surrounding silicates. 
In a second model, \cite{Dahl2010}  considered the possibility of metal-silicate mixing through turbulent entrainment, similar to the model we present here, but their analysis of the structure of the turbulence lead them to conclude that mixing associated with the entrainment process does not proceed to length scales small enough to permit efficient chemical re-equilibration. 
%, though significant large scale mixing is predicted.
Our experiments suggest that mixing does proceed down to the capillary scale at which surface tension balances dynamic pressure fluctuations, and 
our model for the kinetics of equilibration %associated with turbulent entrainment mixing 
predicts fast re-equilibration, % of the metal with the entrained silicates, % since the equilibration distance is predicted to be smaller than the blob diameter, 
implying that  the metal should continuously equilibrate with the entrained silicates once turbulence is well-developed. % (requiring a ... ). 
%We have taken an additional step, developing a model for the kinetics of equilibration associated with turbulent entrainment mixing, 
%%Our model is more optimistic than \cite{Dahl2010} ... and 
%which predicts fast equilibration of the metal with the entrained silicates, % since the equilibration distance is predicted to be smaller than the blob diameter, 
%implying that  the metal should continuously equilibrate with the entrained silicates once turbulence is well-developed. % (requiring a ... ). 
However, remember that fast equilibration between the metal and entrained silicates does not necessarily imply a significant flux of elements from one phase to the other, because this also depends on the amount of metal-silicate mixing. %, as discussed in the introduction and further in section \ref{SectionImplications}. 
Although the entrainment model predicts significantly more mixing than the Rayleigh-Taylor erosion model of \cite{Dahl2010}, chemical re-equilibration remains problematic for the largest impacts (see further discussion in section \ref{SectionImplications}).
}
%and identifying the scales that allow for efficient chemical equilibration. 

\R{
Numerical models of the evolution of a metal blob falling
in molten silicates by \cite{samuel2012} indicate metal fragmentation
occurs through a sequence of events leading to the final
stable drop size, that are quite different from our
experimental results. In our view, a major limitation
of the \cite{samuel2012} study is the assumption of axisymmetry of the
flow, a constraint that inhibits the development
of turbulence.
}

\R{
Regarding the size of the fragments resulting from the metal fragmentation process, models for the maximal stable drop size have been discussed by \cite{Stevenson90}, \cite{Karato1997}, and \cite{Rubie2003}. 
Although we argue for a different scaling for the stable drop size (section \ref{DropSize}), the implications are essentially the same : if the metal phase is fragmented down to the stable drop size, then this size is small enough for ensuring fast chemical re-equilibration between the drops and surrounding silicates. %!!!!
}

\R{\section{Implications for planetary core formation \label{SectionImplications}}}

\begin{figure}[t]
\begin{center}
\includegraphics[width=0.7\linewidth]{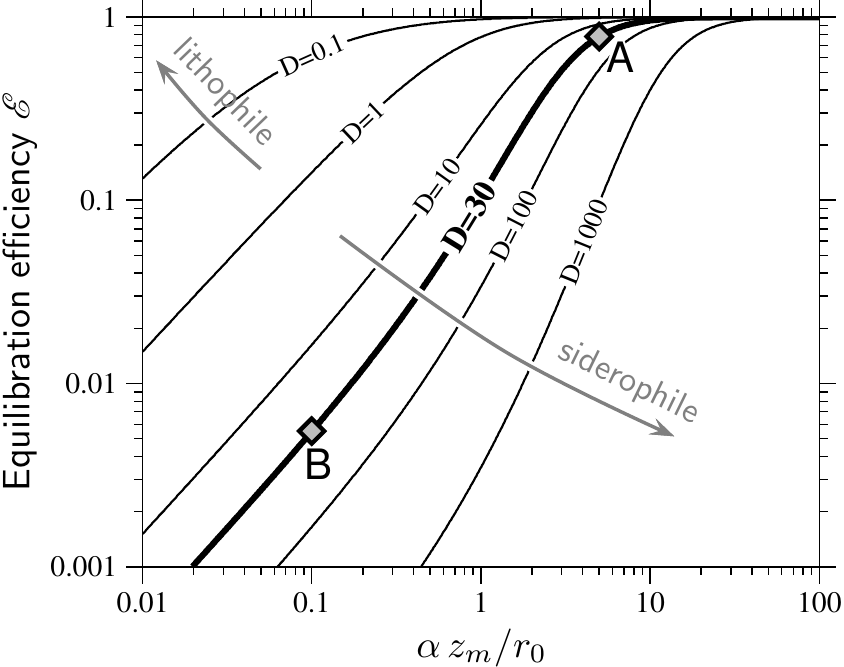}
\end{center}
\vspace*{-0.7cm}
\caption{{Equilibration efficiency} $\mathscr{E}_i$ as a function of $\alpha z_m/r_0$ (where $z_m$ is the depth of the magma ocean) and various values of the partition coefficient $D$, estimated for  metal-silicate mixing in a magma ocean as predicted by the turbulent entrainment model (Eq. \eqref{Phi}).
Point $\mathsf{A}$ corresponds to the case of a metal blob falling through a magma ocean of depth ten times its diameter, with $\alpha=0.25$. 
Point $\mathsf{B}$ corresponds to the case of a giant impact with $r_0= 0.5\, z_m$ and $\alpha=0.05$.
\label{FigEquilibrationEfficiency}
}
\end{figure}

\begin{figure*}[t]
\begin{center}
\includegraphics[width=\linewidth]{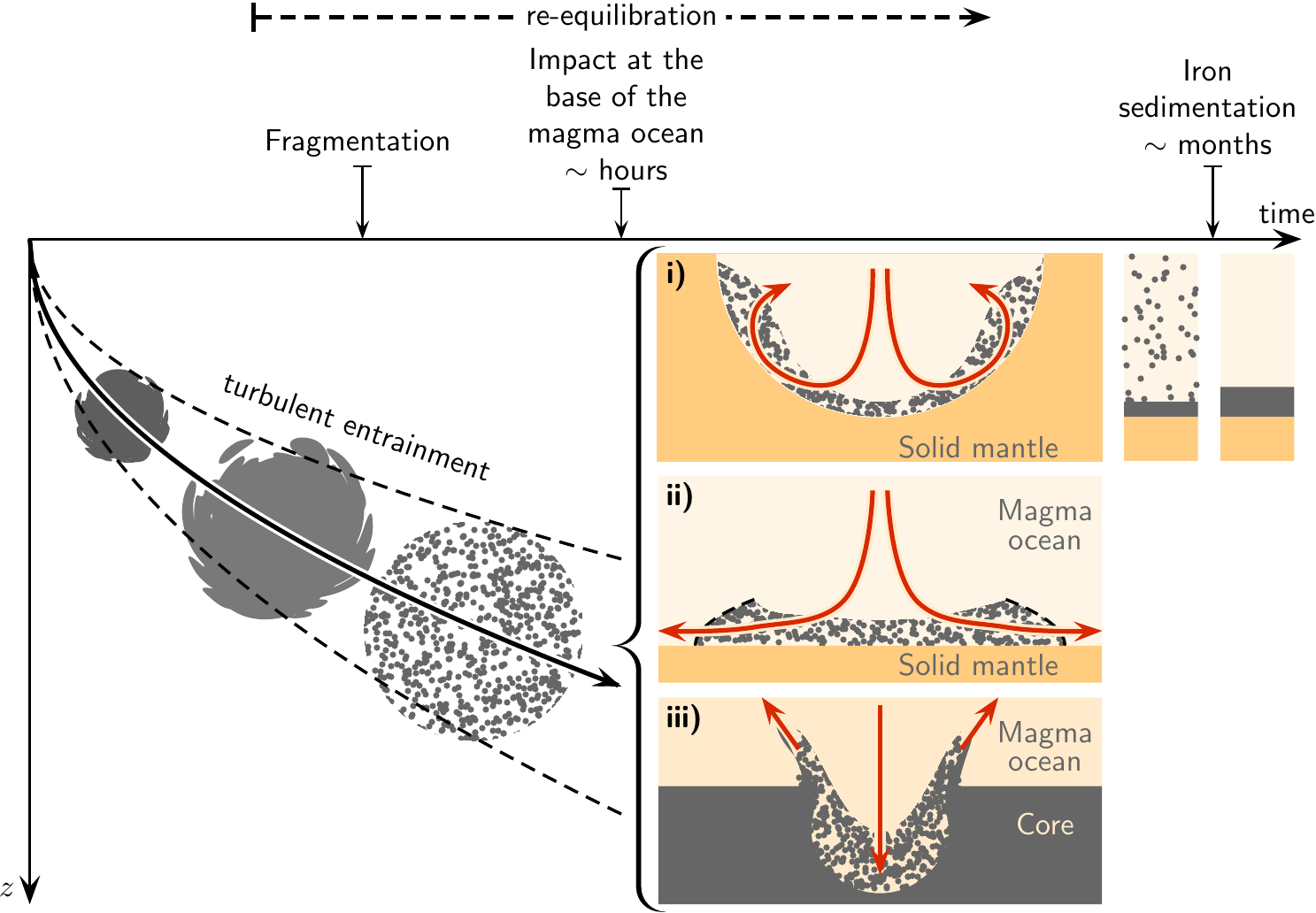}
\caption{{Possible scenarios for metal-silicate mixing and segregation following a large impact involving a previously differentiated impactor.}
The metal is shown in grey, molten silicate in light orange, and solid silicate in dark orange.
The metal phase gradually mixes with the silicates through turbulent entrainment, with efficient chemical equilibration resulting from small-scale mixing.
Additional mixing may be caused by the impact of the metal-silicate mixture at the base of the magma ocean.
\label{Scenarii}
}
\end{center}
\end{figure*}

%\R{This being said ...}

\R{As shown in the introduction section  [see Eq. \eqref{EquilibrationEfficiency}], }
efficient chemical re-equilibration requires that two necessary conditions are met  : (i) that the metal phase is capable of equilibrating with the silicates it has mixed with (\textit{i.e.} that the parameter $k$ in Eq. \eqref{EquilibrationEfficiency} is of order 1), and (ii) that the metal phase equilibrates with a silicate mass  at least a factor $D_i$ larger (\textit{i.e.} that the metal dilution $\Delta \gtrsim D_i$). 

With \R{a velocity} $w$ in the range 0.1-1 km.s$^{-1}$ and \R{diameter} $d > 10$ km, our model predicts that \R{the equilibration distance} $\ell_\mathrm{eq}$ \R{(the distance travelled by the metal phase during the time needed for equilibration)} is always smaller than about $0.6\,d$ (Fig. \ref{Fig_l_eq}).
For example, Eq. \eqref{lEqTimescaleSigma} yields $\ell_\mathrm{eq}\simeq 50$ km for $d=100$ km and $w=100$ m.s$^{-1}$, and $\ell_\mathrm{eq}\simeq 75$ km for $d=1000$ km and $w=1$ km.s$^{-1}$, assuming  $\kappa_c=10^{-8}$ m$^2$.s$^{-1}$, $\rho_s=3500$ kg.m$^{-3}$, $\sigma=1$ J.m$^{-2}$, and $f(\Delta,D_i)= 0.5$.
The corresponding equilibration timescales are $\tau_\mathrm{eq}\simeq 8$ min and $\simeq 75$ s, respectively. 
Since $\ell_\mathrm{eq}$ is smaller than the metal-silicate mixture diameter, and small compared with the typical depth of a magma ocean, the metal phase and the entrained silicate should  readily equilibrate once turbulence is fully developed, which typically requires one advection time $\sim d/w$, or a distance of fall $\sim d$.
%\R{One limitation of the present study is that the time needed for turbulence development is not ...
%In our experiments 
%However, for the largest impacts involving ... similar to the thickness of the magma ocean.
%}
Re-equilibration should be efficient as well once the metal phase is fragmented~: the maximum stable size of the resulting fragments is expected to scale as $d\, W\!e^{-3/5}$  \citep{Kolmogorov1949,Hinze1955,Risso2000}, which predicts submillimeter-to-centimeter size drops, small enough for fast re-equilibration \citep{Karato1997,Rubie2003,Ulvrova2011}.
This suggests that once turbulence is well-developed, most of the metal indeed equilibrates with the surrounding silicates and  $k$ should be close to 1. 
Whether or not metal-silicate equilibration has a significant geochemical fingerprint then depends on the ratio $\Delta/D_i$.  
Assuming that metal-silicate mixing occurs through turbulent entrainment, Fig. \ref{FigEquilibrationEfficiency} shows that the equilibration efficiency $\mathscr{E}_i$, calculated using Eqs. \eqref{EquilibrationEfficiency} and \eqref{Phi} with $k=1$,   depends strongly  on the quantity $\alpha\, z_m/r_0$, where $z_m$ is the depth of the magma ocean.

The above considerations suggest that efficient metal-silicate equilibration should have been the norm for impacts in which the magma ocean  is much deeper than the impactor core diameter.
As an example, Eq. \eqref{Phi} predicts that a molten iron blob  falling through a magma ocean of depth ten times its diameter mixes with about 100 times its mass of silicate, assuming $\alpha=0.25$ (a relevant value here because the large value of $z_m/r_0$ ensures deceleration of the metal phase to subsonic velocity, irrespectively of the initial conditions).
The large value of $z_m/r_0$ also ensures well-developed  turbulence and fast equilibration. 
The resulting Tungsten equilibration efficiency is $\mathscr{E}_W \simeq 0.78$ (point $\mathsf{A}$ in Fig. \ref{FigEquilibrationEfficiency}), assuming $D_W=30$.

The cases of impacts for which $z_m/r_0$ is not much larger than one, which includes the Moon-forming event, are not as clear.
First, it is not obvious that the time needed for the impactor core material to reach the base of the magma ocean would allow enough turbulence to develop and the metal-silicate interfacial area to increase sufficiently for fast equilibration.
Second, \R{and as discussed in section \ref{Limits}}, the effect of compressibility on $\alpha$ may  significantly reduce the entrainment rate,  allowing only a small mass of silicate to mix with the metal.
%Second, \R{and as discussed in section \ref{SectionExperimentalEntrainment}}, the effect of compressibility on $\alpha$ may  significantly reduce the entrainment rate,  allowing only a small mass of silicate to mix with the metal.
Assuming, as for turbulent jets, a fivefold decrease of the entrainment rate due to compressibility, $\alpha=0.25/5=0.05$, the core of an impactor with $\simeq 10$\% the mass of the proto-Earth ($r_0\simeq 0.5\, z_m$) would mix with only about 17~\% its mass of silicate before it reaches the proto-Earth's core, giving $\mathscr{E}_W\simeq 5.5\ 10^{-3}$ (point $\mathsf{B}$ in Fig. \ref{FigEquilibrationEfficiency}). 
However, the actual equilibration efficiency may depend on the details of the impact dynamics.
SPH simulations of the Moon-forming impact suggest that in the likely case of an oblique impact, a fraction of the impactor including most of its core would be sheared past the planet before re-impacting Earth's mantle \citep{Canup2004}.
Some degree of disruption of the impactor core during this process might be sufficient to allow  subsequent metal-silicate equilibration by increasing the value of $\alpha\, z_m/r_0$ for individual blobs.

Lastly, we point out that core-mantle segregation is a complex, multi-step process and  additional equilibration is possible at other stages. 
In particular,  the velocity of the metal-silicate mixture  may easily exceed hundreds of m.s$^{-1}$, implying an energetic  "secondary impact" when it reaches the bottom of the magma ocean, which, as sketched in Fig. \ref{Scenarii}, could cause significant additional metal-silicate mixing \citep{Deguen2011c}. 
(i) In the case of an impact forming its own semi-spherical magma pool, the inertia of the mixture can drive an upward flow, re-suspending iron fragments \citep{Deguen2011c} which, in spite of likely vigorous convection,  sediment out on a timescale similar to the Stokes' sedimentation time  \citep{Martin1988,Lavorel2009}.
(ii) In a pre-existing global magma ocean with a horizontal lower boundary, the metal-silicate mixture will rather spread laterally as a turbulent gravity current - analogous to a pyroclastic flow - 
with possibly significant additional entrainment of molten silicate \citep{Hallworth1993}.
(iii) If the mantle is fully molten, the metal-silicate mixture   directly impacts the proto-Earth's core, with splashing and entrainment of mantle material into the core \citep{Storr1999} providing  additional metal-silicate mixing.  

\vspace*{0.5cm}

\subsection*{Acknowledgement}

We would like to thank H.~J. Melosh, David Rubie, and an anonymous reviewer for helpful comments and suggestions. 
This research was supported by NSF grants EAR-110371  and EAR-1135382 (FESD).

 \appendix

 \section{Equilibration efficiency \label{AppendixEquilibrationEfficiency}}

\paragraph{Definition}
\label{SectionDefinition}

Let $c_m$ and $c_s$ denote the concentrations (in weight \%) of  element $i$ in either the metal or silicate phases, respectively.
The metal and silicate  are fully equilibrated when the two phases have reached thermodynamic equilibrium, for which the equilibrium concentration $c_m^\mathrm{eq}$ and $c_s^\mathrm{eq}$ are linked through the  metal/silicate partition coefficient $D_i$ by $c_m^\mathrm{eq} = D_i\, c_s^\mathrm{eq}$.

Consider a mass $M_m$ of metal, in which we assume that a fraction $k\, M_m$ has been mixed and equilibrated with a mass $M_s$ of silicates.
We define the metal dilution $\Delta$ as the ratio of the mass of equilibrated silicate over the mass of equilibrated metal,
\begin{equation}
\Delta = \frac{M_s}{k M_m}.  \label{DilutionDefinition}
\end{equation}
Given initial values $c_m^0$ and $c_s^0$ of the concentration in the metal and silicate phases, the concentration  in the equilibrated metal $c_m^\mathrm{eq}$ and equilibrated silicate $c_s^\mathrm{eq}$ are found from mass conservation, 
\begin{equation}
c_m^\mathrm{eq} + \Delta\, c_s^\mathrm{eq} = c_m^0 + \Delta\, c_s^0,
\end{equation}
which, together with the assumption of thermodynamic equilibrium, $c_m^\mathrm{eq} = D_i\, c_s^\mathrm{eq}$, gives
\begin{equation}
 c_m^\mathrm{eq}  =  \frac{ c_m^0 + \Delta\, c_s^0}{1  + \Delta /D_i},\quad c_s^\mathrm{eq} = \frac{ c_m^0 + \Delta\, c_s^0}{D_i + \Delta}. \label{Cms_eq}
\end{equation}

The net mass exchange $\mathcal{M}_i$ of element $i$ between the metal and silicate phases can be written as 
\begin{align}
\mathcal{M}_i &= k\,M_m |c_m^\mathrm{eq} - c_m^0| = M_s |c_s^\mathrm{eq} - c_s^0|  \\
                    &= k\,M_m \frac{|c_m^0 - D_i c_s^0|}{1+D_i/\Delta}.  \label{MassExchange}
\end{align}
$\mathcal{M}_i$ reaches a maximum value  $\mathcal{M}_i^\mathrm{max} = M_m |c_m^0 - D_i c_s^0|$ when  all the metal phase is equilibrated ($k=1$) and is infinitally diluted in the silicate phase  ($\Delta \rightarrow \infty$).
We thus define the equilibration efficiency $\mathscr{E}_i$ of element $i$ as  the actual mass exchange $\mathcal{M}_i$ normalized by the maximum possible mass exchange $\mathcal{M}_i^\mathrm{max}$.
From Eq. \eqref{MassExchange} and the value of $\mathcal{M}_i^\mathrm{max}$, the equilibration efficiency is found to be
\begin{equation}
\mathscr{E}_i = \frac{k}{1+D_i/\Delta}, \label{Efficiency}
\end{equation}
which reduces to $k$ when $\Delta/D_i \gg 1$, the limit that is usually assumed in  continuous accretion models  \citep[\textit{e.g.}][]{Rudge2010}.

As shown by Eq. \eqref{Efficiency}, the equilibration efficiency $\mathscr{E}_i$  depends critically on the ratio $\Delta/D_i$, and is small, even when $k=1$, if $\Delta$ is small compared to $D_i$.
Efficient re-equilibration requires  the metal dilution to be similar to or larger than the partition coefficient of the element considered.
For Tungsten, which has $D_W \simeq 30$, efficient re-equilibration thus requires that the metal re-equilibrates with at least $30$ times its mass of silicate.

\paragraph{Use of $\mathscr{E}_i$ in geochemical models}

We demonstrate here that   geochemical models assuming partial equilibration of  the metal phase  but infinite dilution can be  generalized by  using the equilibration efficiency $\mathscr{E}_i$ in place of $k$.
We consider  the case of continuous accretion, according to the formulation of  \cite{Rudge2010} (see their Supplementary Information). 
Discontinuous accretion  can be treated in the same way.

We note $c_m(t)$ and $c_s(t)$ the concentration in Earth's mantle and core at time $t$, and  $c_m^\mathrm{imp}(t)$ and $c_s^\mathrm{imp}(t)$ the composition of the metal and silicate phase of the impacting bodies.
The mass of the Earth is denoted by $M(t)$, and, using $F$ for the mass fraction of metal in the Earth (assumed constant), then the masses of the core and mantle are $F M(t)$ and $(1-F) M(t)$, respectively.
We assume for simplicity that all impactors have the same metal mass fraction $F$.

Conservation of  mass of element $i$ in Earth's core implies that
\begin{equation}
\frac{d}{dt} \left[ F M c_m \right] = \underbrace{(1-k) F c_m^\mathrm{imp} \frac{dM}{dt}}_\text{Flux of non-equilibrated metal} + \underbrace{k F c_m^\mathrm{eq} \frac{dM}{dt}}_\text{Flux of equilibrated metal}   \label{MassConservationCore_1}
\end{equation}
where $c_m^\mathrm{eq}$ is the concentration in the re-equilibrated fraction of the impactor core.
One complication  is that the metal of the impactor may equilibrate with silicates from both the impactor mantle and Earth's mantle, in unknown proportion.
If $\tilde c_s$ denotes the mean composition of the equilibrated silicate,  Eq. \eqref{Cms_eq} yields
\begin{equation}
 c_m^\mathrm{eq}  =  \frac{ c_m^\mathrm{imp} + \Delta\, \tilde c_s}{1  + \Delta /D_i}.  \label{Cms_eq_2}
\end{equation}
For siderophile elements such as Tungsten, $\tilde c_s$ can be  approximated by $c_s(t)$.
As discussed above in  \ref{SectionDefinition}, the effect of re-equilibration is significant only if the metal re-equilibrates with a mass of silicates about $D_i$ times larger (\textit{e.g.} about 30 times larger for Tunsten).
Since the mass of the impactor mantle is only about twice the mass of its core, efficient re-equilibration of siderophile elements requires that the impactor metal equilibrates with a  mass of Earth's mantle significantly larger than the impactor's mantle.
This implies that, in cases where equilibration is efficient, the mean concentration of the equilibrated silicate is close to $c_s(t)$.
The approximation $\tilde c_s \simeq c_s(t)$ is not valid if the equilibration efficiency is small, but in that situation it has little effect on the results.

Substituting Eq. \eqref{Cms_eq_2} into Eq. \eqref{MassConservationCore_1} yields the following equation for the compositional evolution of the core :
\begin{equation}
\frac{d}{dt} \left( M c_m \right) = \left[ \mathscr{E}_i  D_i \tilde c_s + (1-\mathscr{E}_i)  c_m^\mathrm{imp}  \right] \frac{dM}{dt},   \label{MassConservationCore_2}
\end{equation}
while conservation of element $i$ in the mantle yields the following equation for the mantle :
\begin{equation}
\frac{d}{dt} \left(M c_s \right) =  \left[ c_s^\mathrm{imp} +  \mathscr{E}_i \frac{F}{1-F} (c_m^\mathrm{imp} - D_i \tilde c_s) \right] \frac{dM}{dt}.  \label{MassConservationMantle_2}
\end{equation}
If $\tilde c_s$ is taken to be equal to $c_s(t)$, equations \eqref{MassConservationCore_2} and \eqref{MassConservationMantle_2} are  the same as  used by  \cite{Rudge2010} for stable species if $\mathscr{E}_i$ is substituted for $k$ (see their equations A.3 and A.4 in the Supplementary Information).
The equivalence also holds if radioactive or radiogenic species are considered (see the Supplementary Information of  \cite{Rudge2010} for a detailed derivation of the relevant equations).
Results of previous accretion models, including the bounds on Earth's accretion derived by  \cite{Rudge2010} from Hf-W and U-Pb systematics, can therefore be generalized to include the effect of finite dilution by using  $\mathscr{E}_i$ in place of $k$.

\paragraph{Implications}

Previous studies  \citep{Kleine2004,Nimmo2010,Rudge2010} have shown that Hf-W systematics can be used to infer a lower bound for the mean degree of re-equilibration during Earth's accretion.
Assuming infinite dilution of the metal phase,  \citep{Rudge2010} found that Hf-W systematics constrains the fraction of equilibrated metal $k$ to be larger than about 0.36 on average during Earth's accretion.
If finite metal dilution is considered, the implication is that $\mathscr{E}_W\geq \mathscr{E}_W^\mathrm{min} =0.36$, which requires that $k>0.36$ and, assuming $D_W\simeq 30$, $\Delta \geq\Delta^\mathrm{min} = {D_W}/({1/\mathscr{E}_W^\mathrm{min} - 1})\simeq 17$. 

A possibly important implication for modeling the abundance of siderophile elements in the mantle is that the equilibration efficiency $\mathscr{E}_i$ is element-dependent.
One  consequence is that constraints on the equilibration efficiency  from Hf-W systematics do not apply directly to other elements.
The equilibration efficiency of an element $i$ with partition coefficient $D_i$ differs from the Tungsten equilibration efficiency $\mathscr{E}_W$ according to
\begin{equation}
\mathscr{E}_i  = g(D_W,D_i,\Delta)\, \mathscr{E}_W,  \label{EfficiencyRelation}
\end{equation}
where 
\begin{equation*}
g(D_W,D_i,\Delta) = \frac{1+D_W/\Delta}{1+D_i/\Delta}.
\end{equation*}
In Eq. \eqref{EfficiencyRelation}, the function $g$ is an increasing function of  $\Delta$ if $D_i>D_W$, and a decreasing function of $\Delta$ if $D_i<D_W$.
Thus the lower bounds on $k$ and $\Delta$ deduced from Hf-W systematics imply the following lower bound on the equilibration efficiency of an element $i$~:
\begin{equation}
\mathscr{E}_i \geq \mathscr{E}_i^\mathrm{min} =
\begin{cases}
\frac{1+D_W/\Delta^\mathrm{min}}{1+D_i/\Delta^\mathrm{min}} \mathscr{E}_W^\mathrm{min}  &\text{if}\ D_i \geq D_W,\\
\mathscr{E}_W^\mathrm{min} &\text{if}\ D_i \leq D_W.
\end{cases}
\end{equation}
The constraint on the equilibration efficiency  becomes  weaker  for \R{elements that are more siderophile}. For example, the lower bound on the equilibration efficiency is $\mathscr{E}_i^\mathrm{min}\simeq 0.14$ for an element with $D_i = 100$, and only $\mathscr{E}_i^\mathrm{min}\simeq 0.017$ for an element with $D_i = 10^3$.
Thus low equilibration efficiency should be considered  when modeling the  core/mantle partitioning of  highly siderophile elements  \citep[\textit{e.g.}][]{Wood2006,Corgne2008}.

\section{Turbulent entrainment model \label{AppendixTurbulentEntrainment}}

\paragraph{Integral relationships}

We consider a buoyant spherical mass of initial radius $r_0$ and density $\rho_m=\rho_s+\Delta\rho$ released  with an initial (downward) velocity $w_0$ in a fluid of density $\rho_s$.
Owing to entrainment, the mean density of the metal-silicate mixture evolves with time according to
\begin{equation}
\bar \rho(t) = \rho_s + (\rho_m-\rho_s) \phi = \rho_s \left[ 1 + \frac{\Delta \rho}{\rho_s}	 \phi  \right],  \label{MeanDensity}
\end{equation}
where $\phi = r_0^3/r^3$ is the metal phase volume fraction.
The buoyancy of the metal-silicate mixture,  
\begin{equation}
B = g \frac{\bar\rho-\rho_s}{\rho_s} V = g\, \frac{\Delta\rho}{\rho_s}\, \phi\, V, 
\end{equation}
is conserved  in absence of density stratification in the ambient fluid. 
Here $V$ is the volume of the turbulent fluid and $r$ is  its mean radius. 

We adopt the standard entrainment assumption of  \citep{Morton1956} for which the local inward entrainment velocity $u_e$ is proportional to the magnitude of the mean vertical velocity $w$ of the mixture,
\begin{equation}
u_e = \alpha\ |w|,
\end{equation}
where $\alpha$ is the entrainment coefficient.
With this assumption, the equation of conservation of mass becomes
\begin{equation}
\frac{4 \pi}{3}\frac{d (\bar \rho r^3)}{dt} =  4 \pi r^2 \rho_s \alpha |w|,   \label{mass_conservation1} 
\end{equation}
while conservation of momentum becomes  \citep[\textit{e.g.}][]{Bush2003}
\begin{equation}
\frac{4 \pi}{3}\frac{d }{dt}\left[(\bar \rho + K \rho_s) r^3 w\right] = \rho_s B - \frac{1}{2} C_d \rho_s \pi r^2 w^2 .  \label{momentum_conservation1}   
\end{equation}
Here $K$ is the coefficient of added mass, which accounts for the momentum imparted to the surrounding fluid  \citep{Escudier1973}. 
The second term on the right hand side of  equation \eqref{momentum_conservation1} is the hydrodynamic drag $F_d$, with $C_d$ the drag coefficient.

Using Eq. \eqref{MeanDensity} to write $\bar\rho$ as a function of $\phi$, Eqs. \ref{mass_conservation1} and \ref{momentum_conservation1} become
\begin{align}
\frac{d r }{dt} &=   \alpha |w|, \label{mass_conservation2} \\   
\left[ (1 + K ) r^3  + \frac{\Delta\rho}{\rho_s}\, r_0^3\right]\frac{d w}{dt}   &=  g\, \frac{\Delta\rho}{\rho_s}\, r_0^3 - 3 \alpha \left[1 + K + \frac{C_d}{8 \alpha}  \right] r^2 w^2. \label{momentum_conservation2}
\end{align}
Noting that $w=dz/dt$, Eq. \eqref{mass_conservation2} implies that $dr/dz=\alpha$. 

We now non-dimensionalize lengths by $r_0$, time by $\left[\rho_s\,r_0/(\Delta\rho\,g)\right]^{1/2}$, and the velocity by $\left(r_0\, g\,{\Delta\rho}/{\rho_s}\right)^{1/2}$. 
In non-dimensional form,  equations \eqref{mass_conservation2}-\eqref{momentum_conservation2} then become
\begin{align}
\frac{d \tilde{r} }{d\tilde{t}} &=  \alpha |w|, \label{mass_conservation3} \\
\left[ (1 + K ) \tilde{r}^3  + \frac{\Delta\rho}{\rho_s}\right]\frac{d \tilde w}{d\tilde t}   &=  1 - 3 \alpha \left[1 + K + \frac{C_d}{8 \alpha}  \right] \tilde{r}^2 \tilde w^2. \label{momentum_conservation3}
\end{align}
where the tilde ('$\sim$') denotes non-dimensional variables. The initial conditions are
\begin{equation}
\tilde{r}=1,\ \tilde{z}=0,\ \text{and} \ \tilde{w} = \frac{w_0}{\left(r_0\, g\,\frac{\Delta\rho}{\rho_s}\right)^{1/2}} \quad \text{at} \quad \tilde{t}=0.  \label{InitialConditions}
\end{equation}
In Fig. \ref{fig_r_z}, we use a least-square inversion procedure to find the values of $\alpha$, $K$ and $C_d$ for which the model described by Eqs. (\ref{mass_conservation3}-\ref{InitialConditions}) best fits our experimental data on the position of the center of mass $\tilde z$ and radius of the mixture  $\tilde r$ as a function of time.

\paragraph{Analytical solutions}

Using $d\tilde{r}/d\tilde t=\alpha\, d\tilde z/d\tilde t = \alpha\, \tilde w$, Eq. \eqref{momentum_conservation3} can be re-written as
\begin{equation}
\left[ (1+K) \tilde{r}^3 + \frac{\Delta\rho}{\rho_s}  \right] \frac{\alpha}{2}  \frac{d \tilde w^2}{dr} = 1 - 3 \alpha \left[ 1+ K +  \frac{C_d}{8 \alpha} \right] \tilde{r}^2   \tilde w^2 \label{Eq_w2},
\end{equation}
the solution of which is
\begin{equation}
\tilde w^2 =   \frac{2}{\alpha }\int_1^{\tilde{r}}\frac{ \left( \frac{\Delta\rho}{\rho_s} + (1+K) x^3 \right)^{\gamma-1} }{ \left(  \frac{\Delta\rho}{\rho_s} + (1+K)\tilde{r}^3 \right)^{\gamma}}dx   + \left( \frac{  \frac{\Delta\rho}{\rho_s}+1+ K  }{ \frac{\Delta\rho}{\rho_s} + (1+K)\tilde{r}^3 }\right)^{\gamma}\tilde w_0^2 ,   \label{Solution_w2}
\end{equation}
where
\begin{equation}
\gamma =  2+ \frac{C_d}{4(1+K) \alpha} = \frac{2}{1+K}\left( 1 + K + \frac{C_d}{8 \alpha} \right).
\end{equation}
The integral on the RHS of Eq. \eqref{Solution_w2} can be calculated analytically  if $C_d=0$, or if ${\Delta\rho}/{\rho_s}\rightarrow 0$ (for arbitrary $K$ and $C_d$).

The solution \eqref{Solution_w2} has a large-$z$ asymptote given by 
\begin{equation}
\tilde w = \left[2\left( 1+ K + \frac{3}{16} \frac{C_d}{\alpha}\right) \alpha^3  \right]^{-1/2} \frac{1}{\tilde z}  \label{w_SelfSimilar},
\end{equation}
which corresponds to the self-similar regime of a turbulent thermal, consistent with the form given in Eq. (2) of the paper.
Once integrated, Eq. \eqref{w_SelfSimilar} yields
\begin{equation}
\tilde z^2 =  \left[ \left(1+ K + \frac{3}{16} \frac{C_d}{\alpha}\right) \frac{\alpha^3}{2}  \right]^{-1/2} \tilde t.
\end{equation}
$K$ and $C_d$ act in exactly the same way in the self-similar regime. Furthermore, $3/(16\alpha)\sim1$ if $\alpha \simeq 0.25$, which implies that $K$ and $C_d$ have a quantitatively similar effect.

\begin{table}[ht]%[tdp]
\caption{Symbols used in the main text and appendices. \label{Symbols}}
%\hspace*{-3cm}\begin{minipage}{1.5\linewidth}
%\footnotesize
\scriptsize 
%\tiny 
%\begin{center}
\begin{tabular}{ll}
\toprule
\multicolumn{2}{l}{Latin symbols} \\
\midrule
$B$		& Buoyancy of the metal-silicate mixture \\
$c_{m,s}$	& Mean concentration  (wt. \%) of element $i$ in the metal (m)\\ & or silicate (s) phase \\
%$c_s$	& Mean concentration in weight \% of element $i$ in the silicate phase \\
$c_{m,s}^\mathrm{imp}$ & Mean concentration  (wt. \%) of element $i$ in the metal (m)\\ & or silicate (s) phase of the impactor \\
%$c_s^\mathrm{imp}$ &  Mean concentration in weight \% of element $i$ in the silicate phase of the impactor  \\
$c_{m,s}^\mathrm{int}$ & Concentration of element $i$, in the metal (m)\\& or silicate (s) phase at the metal-silicate interface \\
%$c_s^\mathrm{int}$ &  Concentration in weight \% of element $i$ in the silicate phase at the metal-silicate interface  \\
$\Delta c_{m,s}$	& Composition difference across the  boundary layer,\\ &  in the metal (m) or silicate (s) phase	\\
$C_d$	&	Drag coefficient \\
$d$		& Diameter of the metal-silicate mixture \\
$d_\mathrm{max}$	& Maximum stable drop diameter \\
$D$		& Fractal dimension \\
$D_i$	&  Metal/silicate partition coefficient \\
$\mathscr{E}_i$	& Equilibration efficiency \\
$F$	&  Core mass fraction\\
$F\!_c$	& Compositional flux across the metal-silicate interface \\
$g$	& Acceleration of gravity \\
$k$		& Mass fraction of equilibrated metal \\
$K$		& Coefficient of added mass \\
$\ell^*$	& Cut-off length scale \\
$\ell_K$	& Kolmogorov scale \\
$\ell_\sigma$	& Turbulent capillary scale	\\
$\ell_\mathrm{eq}$	& Equilibration distance \\
$\mathcal{M}_i$	& Mass exchange of element $i$ between metal and silicates \\
$\mathcal{M}_i^\mathrm{max}$	& Maximum possible value of $\mathcal{M}_i$ \\
$M_{m,s}$	& Mass of metal (m) or silicate (s) \\
$M$	& Mass of the Earth at time $t$ \\
$r$		& Radius of the metal-silicate mixture \\
%\end{tabular}
%\begin{tabular}{ll}
$S_\ell$	& Area of the metal-silicate mixture measured at scale $\ell$  \\ 
$S_T$	& True area of the metal-silicate mixture  \\ 
$u_\ell$	& Turbulent velocity fluctuation at scale $\ell$	\\
$V$	& Volume of the metal-silicate mixture \\
$w$		& Vertical velocity of the metal-silicate mixture \\
\midrule
\multicolumn{2}{l}{Greek symbols} \\
\midrule
$\alpha$	& Entrainment coefficient \\
$\gamma_{m/s}$	& $\Delta c_{m}/\Delta c_{s}$ \\
$\delta_{m,s}$	& Compositional boundary layer thickness in the metal (m)\\ & or silicate (s) phase \\
$\Delta$	& Metal dilution, \textit{i.e.} the ratio of the mass of equilibrated silicate\\ & over the mass of equilibrated metal \\
$\epsilon$		& Dissipation rate \\
$\eta_{m,s}$	& Dynamic viscosity in the metal (m) or silicate (s) phase	\\
$\kappa_{m,s}$	& Compositional diffusivity in the metal (m) or silicate (s) phase	\\
$\nu_{m,s}$	& Kinematic viscosity in the metal (m) or silicate (s) phase	\\
$\rho_{m,s}$ & Density of the metal (m) or silicate (s) phase  \\
$\bar\rho$ & Mean density of the metal-silicate mixture  \\
$\Delta \rho$	& Density contrast $\rho_m-\rho_s$	\\
$\sigma$		& Metal-silicate interfacial tension \\
$\tau_\mathrm{eq}$	& Equilibration timescale \\
$\phi$	& Metal mass fraction in the metal-silicate mixture \\
\midrule
\multicolumn{2}{l}{Dimensionless numbers} \\
\midrule
$Bo$	& Bond number, $\Delta \rho g d^2 /\sigma$ \\
$\mathsf{H}$	& Viscosity ratio, $\eta_m/\eta_s$ \\
$\mathsf{P}$  & Density ratio, $\rho_m/\rho_s$ \\
$Pe$	& Compositional P\'eclet number, $w\, d/\kappa$\\
$Re$	& Reynolds number, $w\, d/\nu$\\
$S\!c$		& Schmidt number, $\nu/\kappa$ \\
$W\!e$	& Weber number, $\rho_m w^2 d/\sigma$\\
\bottomrule
\end{tabular}
%\end{center}
%\end{minipage}
\label{Notations}
\end{table}%


%% References
%% Usage of  \cite is as follows:
%%    \citep{key}          ==>>  [#]
%%    \cite[chap. 2]{key} ==>>  [#, chap. 2]
%%    \citet{key}         ==>>  Author [#]


\bibliographystyle{model2-names.bst}
%\bibliography{/Users/rdeguen/Boulot/biblio.bib}



\end{document}